\input epsf.tex

\documentclass[11pt]{article}
\usepackage{psfrag}
\usepackage{graphicx}

\newcommand{\bmat}{\left(\begin{array}}
\newcommand{\emat}{\end{array}\right)}

\def\yzero{\smash{\hbox{$y\kern-4pt\raise1pt\hbox{${}^\circ$}$}}}

\def\beq{\begin{equation}}
\def\eeq{\end{equation}}
\def\beqa{\begin{eqnarray}}
\def\eeqa{\end{eqnarray}}

\def\-{\hphantom{-}}
\def\ov{\overline}

\def\s2{\frac{1}{2}}

\def\beq{\begin{equation}}
\def\eeq{\end{equation}}
\def\beqa{\begin{eqnarray}}
\def\eeqa{\end{eqnarray}}

\def\Tr{{\rm Tr \,}}

\def\IF{\relax{\rm I\kern-.18em F}}
\def\II{\relax{\rm I\kern-.18em I}}

\def\IC{\bf C}
\def\IZ{\bf Z}
\def\IR{\bf R}

\def\IS{\bf S}

\def\IT{\bf T}

\def\z2z2{$\IC^3/(\IZ_2\times\IZ_2)$}

\def\id{{\bf 1}}

\def\Dsl{\,\raise.15ex\hbox{/}\mkern-13.5mu D} 

\newcommand{\drawsquare}[2]{\hbox{%
\rule{#2pt}{#1pt}\hskip-#2pt
\rule{#1pt}{#2pt}\hskip-#1pt
\rule[#1pt]{#1pt}{#2pt}}\rule[#1pt]{#2pt}{#2pt}\hskip-#2pt
\rule{#2pt}{#1pt}}

\newcommand{\fund}{\raisebox{-.5pt}{\drawsquare{6.5}{0.4}}}
\newcommand{\antifund}{\overline{\fund}}

\begin{document}

\makeatletter \@addtoreset{equation}{section} \makeatother
\renewcommand{\theequation}{\thesection.\arabic{equation}}

\pagestyle{empty}
\vspace*{.5in}
\rightline{CERN-PH-TH/2006-091}
\rightline{IFT-UAM/CSIC-06-22}
\rightline{\tt hep-th/0605166}
\vspace{1.5cm}

\begin{center}
\LARGE{\bf Local models of Gauge Mediated Supersymmetry Breaking in 
String Theory} \\[10mm]

\medskip

\large{I\~naki Garc\'{\i}a-Etxebarria$^\dagger$, Fouad Saad$^\dagger$,
Angel
M. Uranga$^\ddagger$}
\\
{\normalsize {\em $^\dagger$ Instituto de F\'{\i}sica Te\'orica, C-XVI \\
Universidad Aut\'onoma de Madrid \\
Cantoblanco, 28049 Madrid, Spain \\
$^\ddagger$ TH Unit, CERN, \\
CH-1211 Geneve 23, Switzerland \\
{\tt innaki.garcia@uam.es, fouad.saad@uam.es, angel.uranga@cern.ch,
angel.uranga@uam.es}
\\[2mm]}}

\end{center}

\smallskip

\begin{center}
\begin{minipage}[h]{14.5cm}
{\small
We describe local Calabi-Yau geometries with two isolated singularities at 
which systems of D3- and D7-branes are located, leading to chiral sectors 
corresponding to a semi-realistic visible sector and a hidden sector with 
dynamical supersymmetry breaking. We provide explicit models with a 
3-family MSSM-like visible sector, and a hidden sector breaking 
supersymmetry at a meta-stable minimum. For singularities separated by a 
distance smaller than the string scale, this construction leads to a 
simple realization of gauge mediated supersymmetry breaking in string theory. 
The models are simple enough to allow the explicit computation
of the massive messenger sector, using dimer techniques for branes at 
singularities. The local character of the configurations makes manifest 
the UV insensitivity of the supersymmetry breaking mediation.
}
\end{minipage}
\end{center}

\newpage

\setcounter{page}{1} \pagestyle{plain}
\renewcommand{\thefootnote}{\arabic{footnote}}
\setcounter{footnote}{0}

\section{Introduction}

\label{intro}

The study of low energy supersymmetry and supersymmetry breaking are the 
main driving forces in present research in physics beyond the Standard 
Model. Hence, their description and understanding in terms of an 
underlying theory is highly desirable. 

String theory implements 
supersymmetry at high energies automatically, and has enough richness to 
allow for mechanisms of supersymmetry breaking, and its mediation to the 
Standard Model sector. A nice scenario is supersymmetry breaking in 
a hidden sector with gravity mediation, and a particularly nice  
realization is in flux compactifications (see 
\cite{Polchinski:1995sm,Becker:1996gj,Dasgupta:1999ss,Giddings:2001yu}, etc), 
with the moduli acting as hidden sector. In this particular setup, 
techniques to obtain the soft terms have been devised
\cite{Grana:2002nq,Camara:2003ku,Grana:2003ek,Camara:2004jj,Lust:2004fi,Lust:2004dn,Font:2004cx,Lust:2005bd,Gomis:2005wc}, 
(some exploiting earlier model-independent approaches 
\cite{Kaplunovsky:1993rd,Brignole:1993dj,Brignole:1995fb}).

Gauge mediated supersymmetry breaking (GMSB) is a purely field theoretical 
mechanism of supersymmetry breaking mediation, insensitive to UV dynamics 
(and hence to gravity). Still it is important to understand its 
realization in a complete theory like string theory. 
This requires the construction of string theory configurations with 
gauge sectors whose (presumably strong) dynamics is under control.
Hence we may expect great benefits from recent developments in the 
understanding of gauge theory dynamics in D-brane setups, mainly 
motivated by the gauge/string correspondence. For instance, the study of 
configurations of D3-branes at singularities, in the presence of 
fractional branes, has led to the realization of large classes of gauge 
theories with strong infrared dynamics, giving rise to interesting phenomena 
like different patterns of confinement \cite{Klebanov:2000hb,Franco:2005fd}
(by the so-called deformation fractional branes), or the removal of the 
supersymmetric vacuum 
\cite{Berenstein:2005xa,Franco:2005zu,Bertolini:2005di}
(by the so-called DSB fractional branes) \footnote{As explained in 
Section \ref{dsb}, the gauge theory on 
such fractional branes has runaway dynamics, rather than a 
non-supersymmetric minimum
\cite{Franco:2005zu} (see also \cite{Intriligator:2005aw,Brini:2006ej}). 
However, it has been recently shown in 
\cite{Franco:2006es} that, in a generalization of 
\cite{Intriligator:2006dd}, these theories have long-lived local 
supersymmetry-breaking minima when additional flavours (arising 
from D7-branes) are added.}.

In fact, the latter configurations were explicitly used in
\cite{Diaconescu:2005pc} in the construction of string
compactifications with semi-realistic visible sectors and a sector of
DSB branes. These are the first serious attempts to implement GMSB in
string theory.

In this paper we continue along those lines, improving it in several 
directions. We propose a fairly general framework to discuss models of 
GMSB in string theory. The construction is based on the use of local 
(namely non-compact) configurations, with two sectors of D-branes 
describing the visible and supersymmetry breaking sector, decoupled at 
the massless level, but coupled via a messenger sector
whose mass scale is controlled by the distance between the D-brane 
sectors, which is much smaller than the string scale. In fact, 
it is this latter fact that motivates considering {\em local} 
configurations, since the physics of the mediation is naturally 
insensitive to the global structure of the compactification
\footnote{Of course, in the regime of distance much larger than the 
string scale, the system corresponds to a model of gravity mediation, but 
the latter would be sensitive to the global structure of the 
compactification, hence rendering the local model less useful.}. 
We propose 
explicit realizations of this construction, which is nevertheless quite 
flexible and allows for many generalizations.

Some of the nice features of our proposal and explicit models are:

$\bullet$ Being local, they manifestly show the UV insensitivity of the 
construction.

$\bullet$ As opposed to previous proposals, the computation of the 
spectrum and interactions of the messenger sector can be explicitly 
described.

$\bullet$ The construction is simple and flexible enough to allow for 
many generalizations.

We find that these nice features are an important step in improving models 
of GMSB in string theory. 

Our constructions are based on local Calabi-Yau geometries with two 
isolated singularities, at which sectors of D-branes are located. 
In the construction of the geometries and the determination of the gauge 
sector, we invoke important recent developments on D-branes at 
singularities, especially dimer diagrams (or brane tilings) 
\cite{Hanany:2005ve,Franco:2005rj,Hanany:2005ss,Feng:2005gw,Franco:2006gc,Garcia-Etxebarria:2006aq}, 
which we review in order to make the paper self-contained. The 
location of the D-branes at singularities is a natural way to obtain
chiral 4d $N=1$ gauge theories, rich and flexible enough to allow for 
semi-realistic sectors and sectors with supersymmetry breaking dynamics.
In particular, we can construct examples where the visible sector is an 
MSSM like model with the Standard Model gauge factors and 3-families of 
quarks and leptons, introduced in \cite{Aldazabal:2000sa}, and the 
supersymmetry breaking sector is provided by the flavored $dP_1$ models in 
\cite{Franco:2006es}. Implementation of other concrete models in our 
framework is possible as well, and we mention several possible 
generalizations (in particular, we discuss how to include in our setup 
models with visible sectors based on non-abelian orbifold singularities, like 
$\IC^3/\Delta_{27}$  
\cite{Aldazabal:2000sa,Berenstein:2001nk,Verlinde:2005jr}).

The paper is organized as follows. In Section \ref{background} we provide
background material on dimer diagrams: Section \ref{dimer} describes the 
gauge theories on configurations of D3- and D7-branes at singularities 
using the tools from dimer diagrams. Section \ref{dsb} reviews the 
construction of gauge theories with supersymmetry breaking using 
D-branes. Section \ref{split} describes the construction of local CY 
models with several separated singularities, by using partial resolution. 

In Section \ref{basic} these tools are put to work
in the construction of a simple local CY with two sectors, corresponding to 
D3-branes at two separated singularities. One D3-brane stack describes the 
visible sector (in a toy version given by a 3-family $SU(3)^3$ 
trinification model)
while the other describes the supersymmetry breaking sector (although it 
actually corresponds to a theory with a runaway behaviour). In Section 
\ref{complete} a more complete model is presented, based on the previous 
model with the addition of D7-branes. The visible sector is given by a 
3-family MSSM-like theory, while the hidden sector breaks supersymmetry in 
a local metastable minimum. Such explicit constructions are amenable to 
the study of several phenomenological questions. In Section \ref{possib} 
we sketch other model building possibilities. Finally in Section 
\ref{comments} we present some final remarks. The computation of the 
massive messenger sector in this general class of models is presented in 
Appendix \ref{massive}. 

\section{Background material}
\label{background}

\subsection{D-branes at singularities and dimer diagrams}
\label{dimer}

\subsubsection*{D3-branes at singularities}

Systems of D3-branes at singularities have been under intense study from 
the viewpoint of the AdS/CFT correspondence 
(starting with \cite{Klebanov:1998hh}, see  
\cite{Bertolini:2004xf,Benvenuti:2004dy,Benvenuti:2005ja,Franco:2005sm,Butti:2005sw,Butti:2005vn}
for some recent references) and extensions to related non-conformal systems
(see e.g. 
\cite{Klebanov:2000hb,Franco:2004jz,Herzog:2004tr,Franco:2005fd,Berenstein:2005xa,Franco:2005zu,Bertolini:2005di}).

Another useful viewpoint on these systems is to consider them local 
models of interesting gauge/D-brane dynamics, often illustrating 
properties of more general configurations. In particular, they can be  
regarded as a local description of a global compactification, in a regime 
where the relevant D-branes are close to each other, as compared with the  
compactification scale. This is the viewpoint we take in the present 
paper, in the spirit of \cite{Aldazabal:2000sa}.

A recent useful tool in the study of D3-branes at singularities is 
provided by the dimer diagrams or brane tiling
\cite{Hanany:2005ve,Franco:2005rj,Hanany:2005ss,Feng:2005gw}, 
which we review in this Section.

D3-branes located at a singularity in the transverse space lead to 4d 
gauge theories in their world-volume. For Calabi-Yau singularities, these
theories are $N=1$ and are characterized by a set of gauge factors, chiral
multiplets in bi-fundamental representations, and superpotential interactions
among them. This structure is nicely encoded in the so-called dimer diagrams
(or brane tilings) 
\cite{Hanany:2005ve,Franco:2005rj,Hanany:2005ss,Feng:2005gw,Franco:2006gc}.
They correspond to a periodic tiling of $\IR^2$ or 
equivalently a tiling of the 2-torus $\IT^2$. In order to correspond to a 
gauge theory on D3-branes at a singularity, there are further constraints 
on the tiling. The main one is that the graph should be bi-partite, namely 
the nodes can be colored with two colours (black and white), with edges 
joining nodes of different color. We skip further discussions, and refer 
the reader to e.g. \cite{Franco:2005rj,Hanany:2005ss} for details. 

In this language, each face corresponds to a gauge 
factor, each edge separating two faces corresponds to a chiral multiplet 
in the bi-fundamental representation, and each node corresponds to an 
interaction term in the superpotential, involving the bi-fundamentals 
corresponding to the edges ending on that node \footnote{\label{bivalent}
Bi-valent nodes (i.e. those with just two edges) correspond to mass terms 
for fields which can be eliminated 
by integrating out. The dimer diagram of the resulting theory is obtained 
by simply removing the two bi-valent node and its two edges, and collapsing 
the adjacent nodes.}. Note that the orientation 
on the edges (e.g. from black to white nodes) must be used to define the 
bi-fundamentals. Also, superpotential terms associated to black or white 
nodes have opposite signs. 

\begin{figure}
\begin{center}
\centering
\epsfysize=3.5cm
\leavevmode
\epsfbox{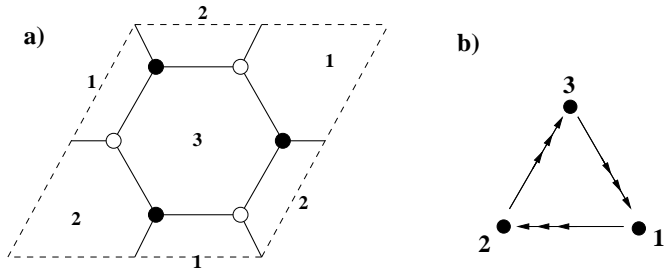}
\end{center}
\caption[]{\small (a) The dimer diagram (as a tiling of the $\IT^2$ upon 
identifying sides of the paralelogram) and (b) the quiver diagram of the 
gauge theory on D3-branes at the $\IC^3/\IZ_3$ singularity.}
\label{dp0dimer}
\end{figure}

One example, corresponding to  D3-branes at the $\IC^3/\IZ_3$ singularity 
(also known as the complex cone over $dP_0$, hence denoted $dP_0$ 
singularity in the following), is shown in Figure \ref{dp0dimer}a. 
The gauge theory corresponding to the dimer
diagram in Figure \ref{dp0dimer}a is described in Figure \ref{dp0dimer}b 
in terms of its quiver diagram, where nodes correspond to gauge factors, 
arrows correspond to chiral multiplets, and the superpotential needs to 
be specified explicitly. In this case we have
\beqa
W & = 
& \Tr \, ( \, X_{12} Y_{23} Z_{31}  \, -\, X_{12} Z_{23} Y_{31} \, +
\, X_{23} Y_{31} Z_{12} \, -\, X_{23} Z_{31} Y_{12} \, + \nonumber \\
& + & X_{31}Y_{12}Z_{23} \, - \, X_{31} Z_{12} Y_{23} \, ) \, \simeq 
\,\epsilon_{ijk}\, \Tr (\, X_{12}^{(i)} X_{23}^{(j)} 
X_{31}^{(k)}\, )
\label{dp0supo}
\eeqa
with obvious notation (in the last expression we have written $X^{(i)}$,
$i=1,2,3$ for $X$, $Y$, $Z$, respectively). Traces in superpotential terms 
will be implicit in what follows.

\begin{figure}
\begin{center}
\centering
\epsfysize=3.5cm
\leavevmode
\epsfbox{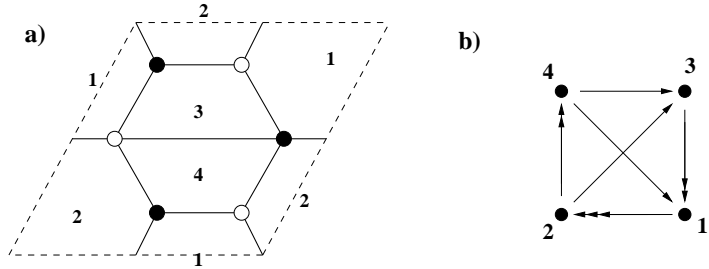}
\end{center}
\caption[]{\small The dimer diagram (a) and quiver diagram (b) of the 
gauge theory on D3-branes at a singularity given by a complex cone over 
$dP_1$ (the $dP_1$ theory, for short).}
\label{dp1dimer}
\end{figure}

For completeness and future use, we show another example of a dimer diagram 
in Figure \ref{dp1dimer}a, corresponding to D3-branes at a singularity 
given by the complex cone over $dP_1$ (in what follows, $dP_1$ singularity 
for short). The corresponding gauge theory (denoted $dP_1$ theory) is 
described by the quiver diagram shown in Figure \ref{dp1dimer}b, with the 
superpotential given by
\beqa
W & = & X_{12} Y_{24} X_{41}\, - \, Y_{12} X_{24} X_{41}\, +\, X_{31} 
Y_{12} X_{23} \, - \nonumber \\
&-& Y_{31}X_{12} X_{23}\, +\, Z_{12} X_{24} X_{43} Y_{31} 
\, -\, Z_{12} Y_{24} X_{43} X_{31}\, \nonumber \\
& \simeq & \epsilon_{ij} X_{12}^{i} X_{24}^j X_{41} \, + \, \epsilon_{ij}
X_{31}^i X_{12}^j X_{23} \, +\, \epsilon_{ij} Z_{12} X_{24}^i X_{43} 
X_{31}^j 
\eeqa
where fields $X^i$, $i=1,2$ denote $X$, $Y$.

Dimer diagrams have been shown to encode the string theory geometry in 
several ways. In the following we provide the most practical description 
for our purposes.

A toric Calabi-Yau geometry is characterized by its web diagram, see 
\cite{Aharony:1997ju,Aharony:1997bh,Leung:1997tw} for a first
application in the physical context and e.g. \cite{Franco:2005fd}
for applications to systems of D3-branes \footnote{The web diagram is dual 
to the toric diagram for the singularities. In particular examples we 
will show figures with the toric diagrams for our geometries for the 
benefit of readers familiar with them; they can however be safely skipped 
by those who are not.}. The web diagram for a toric singularity is given 
by a set of segments in $\IR^2$, carrying $(p,q)$ labels which define 
their orientation \footnote{For a given singularity, the web 
diagram is defined up to an overall $SL(2,\IZ)$ transformation on the 
$(p,q)$ labels.}. Segments join at vertices, with the rule that the 
$(p,q)$ charges of segments at a vertex add up to zero.

\begin{figure}
\begin{center}
\centering
\epsfysize=3.5cm
\leavevmode
\epsfbox{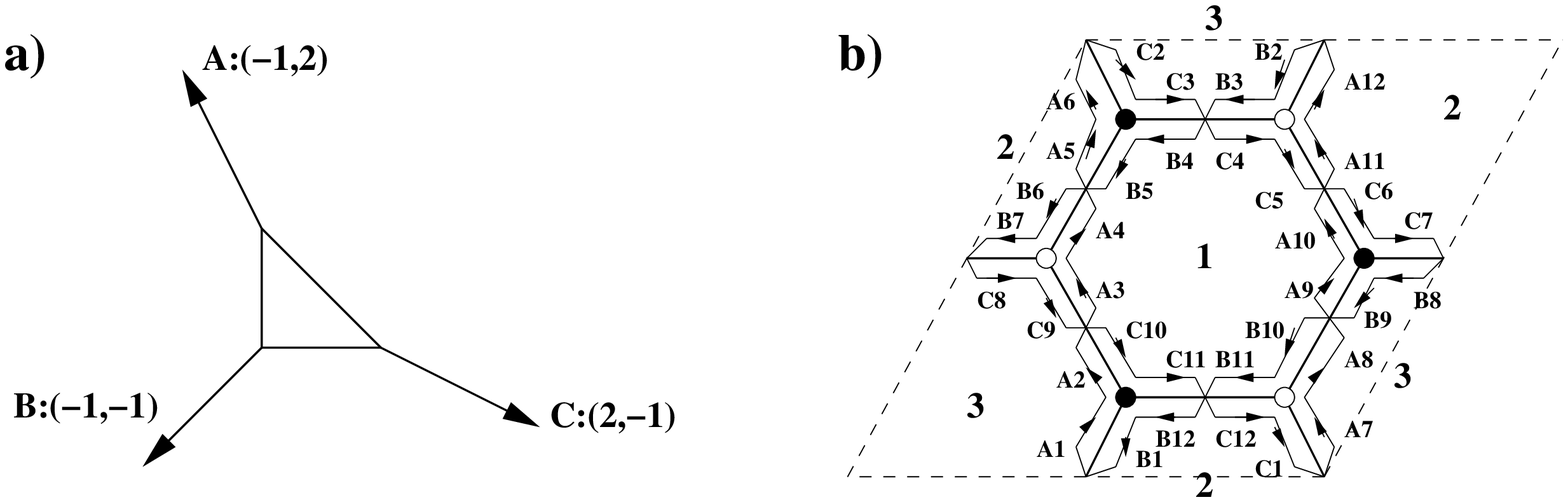}
\end{center}
\caption[]{\small 
(a) Web diagram for the $dP_0$ singularity. For clarity we show the 
geometry for a non-zero size of the internal pieces. (b) Dimer 
diagram and zig-zag paths for the $dP_0$ theory. The $(p,q)$ homology 
class of the path is related to the $(p,q)$ label of an external leg in 
the web diagram of the geometry.} 
\label{dp0zigzag}
\end{figure}

\begin{figure}
\begin{center}
\centering
\epsfysize=3.5cm
\leavevmode
\epsfbox{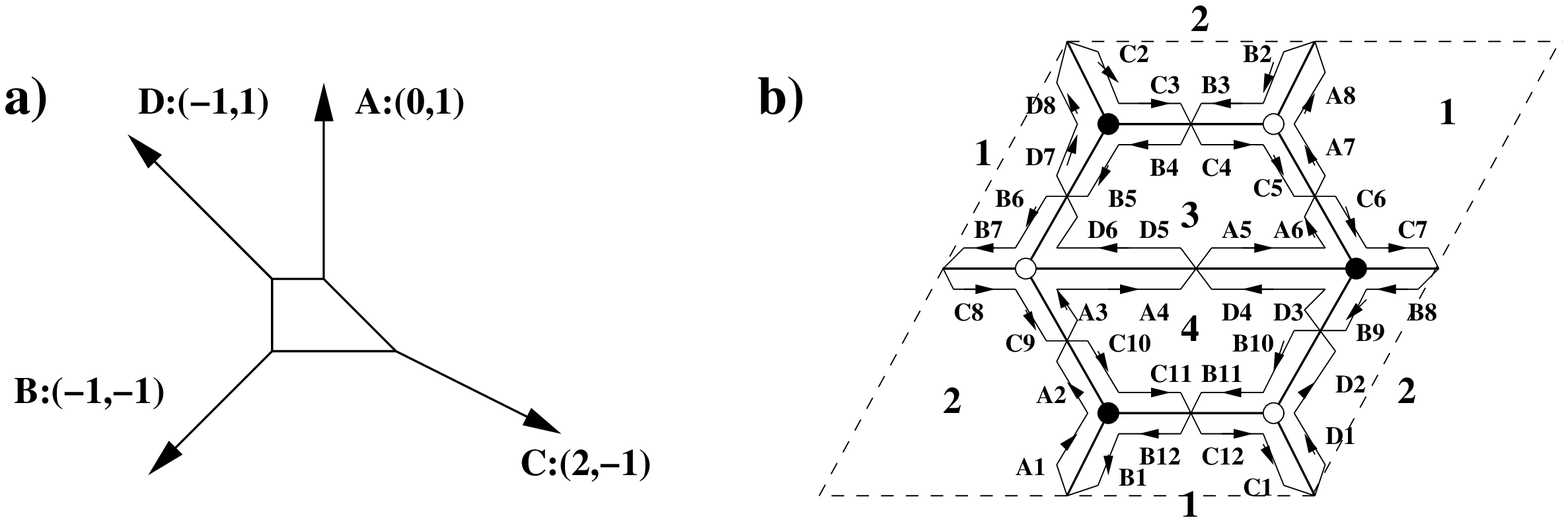}
\end{center}
\caption[]{\small 
(a) Web diagram for the $dP_1$ singularity. For clarity we show the 
geometry for a non-zero size of the internal pieces. (b) Dimer 
diagram and zig-zag paths for the $dP_1$ theory. The $(p,q)$ homology 
class of the path is related to the $(p,q)$ label of an external leg in 
the web diagram of the geometry.} 
\label{dp1zigzag}
\end{figure}

The web diagram for the $\IC^3/\IZ_3$ ($dP_0$) and the $dP_1$ 
singularities are shown in Figures \ref{dp0zigzag}a and \ref{dp1zigzag}a. 
The web diagram can be regarded as describing the
locus where certain $\IS^1$ fibrations in the toric geometry degenerate. 
Skipping the details, this description implies that finite size segments 
and faces correspond to 2-cycles and 4-cycles respectively. External legs 
and non-compact faces correspond to non-compact 2- and 4-cycles.
The structure of the singularity is specified by the
set of $(p,q)$ charges of the external legs, while the sizes of the 
internal finite size pieces simply corresponds to a choice of Kahler moduli.
The singular variety corresponds to shrinking the finite pieces to a point.

The dimer diagram for the gauge theory on D3-branes at a singularity encodes
the $(p,q)$ charges of the external legs in the corresponding web diagram
\cite{Hanany:2005ss,Feng:2005gw}, in its structure of zig-zag paths. A 
zig-zag path is a path made of dimer edges, such that it
turns maximally to the left at e.g. black vertices and maximally to the right
at white vertices. Each zig-zag path defines a closed loop on $\IT^2$, and 
carries a non-trivial $(p,q)$ homology charge. Each zig-zag path corresponds
to an external leg in the web diagram, with the $(p,q)$ label of the leg given
by the $(p,q)$ charge of the path. It is easy to recover the web diagrams
of different singularities from the zig-zag paths of the dimer diagram, as
the reader can check in our examples. The structure of zig-zag paths for 
the $dP_0$ and $dP_1$ dimer diagrams are shown in Figures 
\ref{dp0zigzag}b and \ref{dp1zigzag}b.

We would like to mention a more advanced concept, the mirror Riemann
surface $\Sigma$ and its relation to the dimer. This is useful in the 
derivation of some results, although we will always provide the final 
answers in a language not involving it, so that the reader can safely skip 
them (we refer to \cite{Feng:2005gw} for further details).
The web diagram of a toric singularity can also be regarded as a skeleton 
for a Riemann surface $\Sigma$ with punctures, which is obtained by 
`thickening' the segments to tubes. Punctures in $\Sigma$ 
correspond to external legs in the web diagram. This Riemann surface 
plays a prominent role in the description of the mirror geometry, and
all relevant D-branes are described as wrapped on 1-cycles on it.
In particular, the D3-brane gauge factors correspond to non-trivial 
compact 1-cycles in $\Sigma$. The number of intersections (counted with 
orientation) between two such 1-cycles gives the number of bi-fundamentals 
between the corresponding gauge factors. Finally, the superpotential terms 
correspond to disks in $\Sigma$ bounded by pieces of different 1-cycles. 
Although this picture underlies the derivation of our tools, we
rephrase the results directly in terms of the dimer diagram

\subsubsection*{Adding D7-branes}

For certain applications, it is desirable to introduce D7-branes passing 
through a system of D3-branes at a singularity. Namely, one introduces
D7-branes wrapped on holomorphic 4-cycles of the singular CY. From the 
viewpoint of the 4d gauge theory, this implies the introduction of a set
of flavours for the different D3-brane gauge factors (from the open strings
between the D3- and D7-branes) and interactions (e.g. from 73-33-37 
interactions). Notice that the gauge group on the D7-branes behaves as a
global symmetry from the viewpoint of the 4d gauge theory in this 
non-compact setup.

It turns out that the introduction of such D7-branes can be easily described 
in the language of dimer diagrams, in a manner that allows reading off the
D3-D7 spectrum and interactions. This has been described in appendix B of 
\cite{Franco:2006es}, whose results we briefly review. 
Using the mirror Riemann surface $\Sigma$ mentioned above, 
D7-branes are represented as non-compact 1-cycles in $\Sigma$
that stretch between two punctures. The intersections of the D7-brane 
1-cycle with the 1-cycle corresponding to the D3-branes gives rise to
chiral multiplets in bi-fundamentals of the D3- and D7-brane symmetry groups,
thus providing the D3-D7 spectrum. Finally, disks in $\Sigma$ bounded by
one D7-brane 1-cycle and two D3-brane 1-cycles lead to a cubic 
superpotential term of the form 73-33-37. This more detailed description
underlies our above recipe, which we nevertheless can state directly in 
terms of the dimer diagram.

\begin{figure}
\begin{center}
\epsfxsize=10cm
\hspace*{0in}\vspace*{.2in}
\epsffile{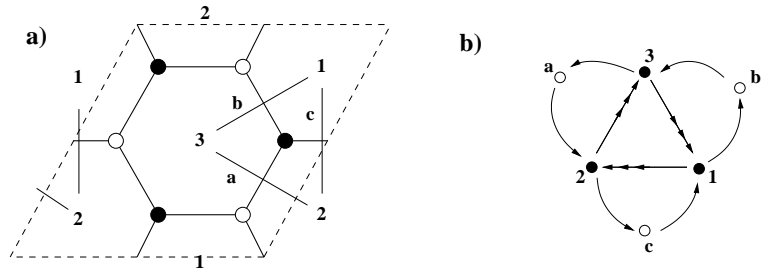}
\caption{\small (a) Extended dimer diagram of the $dP_0$ theory with some 
examples of D7-branes represented as segments across the edges. (b) 
Quiver diagram including D7-branes (represented as white nodes). There 
are 33-37-73 couplings involving the 33 bi-fundamental across which the 
corresponding D7-brane stretches.}
\label{dp0d7s}
\end{center}
\end{figure}

As described in \cite{Franco:2006es}, for each 33 bi-fundamental in the 
D3-brane sector, there exists one kind of D7-brane leading to 37, 73 
chiral multiplets coupling to the 33 state. Hence, a 
simple representation of a D7-brane in the dimer diagram is as a segment 
stretching across an edge, joining the mid-points of adjacent faces. One 
such segment stretching across an edge associated with an 
$(\fund_1,\antifund_2)$ bi-fundamental, gives rise to chiral multiplets in 
the $(\fund_2;\ov{N_{D7}})$ and $(N_{D7};\antifund_1)$, where $N_{D7}$, 
$\ov{N_{D7}}$ represent the D7-brane global symmetries. Heuristically, the 
D7-brane segment touches the faces at its endpoints, leading to the D7-D3 
and D3-D7 sectors according to orientation. There is a 
superpotential coupling 33-37-73 involving these states. The 
representation as a segment facilitates an easy 
identification of the gauge theory matter content and interactions 
corresponding to a system of D3- and D7-branes at singularities. In 
Figure \ref{dp0d7s} we show one particular example of 
this kind of diagram, which we denote extended dimer diagram.
Notice that there are other possible D7-brane choices, namely one for 
each 33 bi-fundamental, and that for different 33 bi-fundamentals with 
the same gauge quantum numbers, the corresponding D7-branes lead to the 
same 37, 73 spectrum, but different 33-37-73 interactions.

An important point is that there are non-trivial consistency conditions
on configurations of D3- and D7-branes at singularities. Concretely, the total
charge of the D-brane system under RR fields living at the singular points
should vanish. Equivalently 
\cite{Leigh:1998hj,Aldazabal:1999nu,Bianchi:2000mi}, the 4d gauge theory 
should be free of non-abelian anomalies \footnote{
\label{uunos} As discussed in 
\cite{Ibanez:1998qp,Morrison:1998cs}, the $U(1)$ mixed
anomalies are canceled by a Green-Schwarz mechanism, and do not require
additional constraints. Also, all anomalous $U(1)$s (plus some 
non-anomalous ones in certain cases) have $B\wedge F$ couplings, which 
gives them a mass of order the string scale. The only linear combination 
of $U(1)$s that generically remains massless is the `diagonal' 
combination $\sum_a \frac{1}{N_a} Q_a$, where $Q_a$ is the $U(1)$ 
generator of the $a^{th}$ gauge factor $U(N_a)$.}. In all our forthcoming 
examples we enforce this property.

\begin{figure}[!hbp]
\centering
\psfrag{QL}{$Q_L$}
\psfrag{UR}{$U_R$}
\psfrag{DR}{$D_R$}
\psfrag{DRb}{$\ov{D_R}$}
\psfrag{L}{$L$}
\psfrag{E}{$E$}
\psfrag{HU}{$H_U$}
\psfrag{HD}{$H_D$}
\includegraphics[scale=0.50]{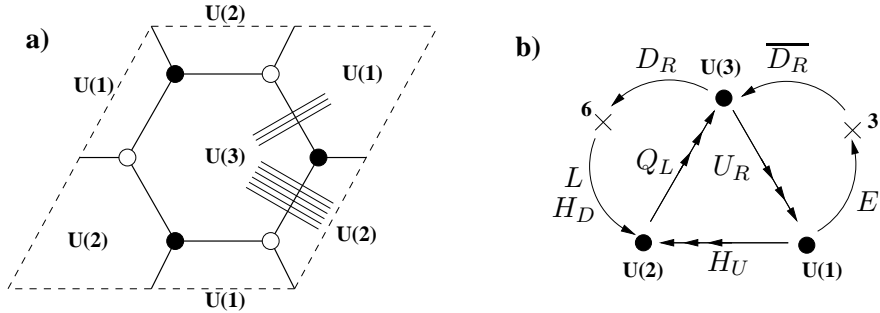}
\caption{\small Dimer diagram (a) and quiver diagram (b) for a 
configuration of D3/D7-branes realizing a gauge theory close to the MSSM.}
\label{dp0mssm}
\end{figure}

One can use these tools to construct interesting gauge theories.
As a particular application to phenomenological model building, it is easy 
to construct configurations leading to MSSM like spectra 
\cite{Aldazabal:2000sa}. In Figure \ref{dp0mssm}a we show an extended 
dimer diagram for a system of D3- and D7-branes at a
$\IC^3/\IZ_3$ singularity studied in \cite{Aldazabal:2000sa}. As can be 
easily read out from the picture, it leads to a $U(3)\times U(2)\times 
U(1)$ gauge group and
3-families of quarks and leptons (plus additional fields, with vector-like
quantum numbers under the Standard Model gauge group). The only massless
$U(1)$ linear combination (in a convenient normalization) is 
$Q_Y=-\frac 12(\frac 13 Q_3 +\frac 12 Q_2 + Q_1)$. This is crucial, since 
it precisely reproduces the correct hypercharges of the matter fields.
In Figure \ref{dp0mssm}b we show the quiver diagram for this gauge theory
\footnote{As discussed before, several 
possible D7-branes can lead to the same D3-D7 spectrum (but different 
interactions). Our dimer diagram is just one of the possible ones leading 
to the same chiral spectrum.}, with arrows labeled by the corresponding 
(Minimal Supersymmetric) Standard Model field. Notice that, in contrast 
with the MSSM, the model contains a triplicated sector of Higgs fields, 
and also that there are three copies of fields, vector-like under the 
D3-brane gauge interactions, with quantum numbers of $D_R$ quarks (and 
conjugates $\ov{D_R}$). See \cite{Aldazabal:2000sa} for further details. 
In later 
sections, we will use this configuration as our (toy) model for the 
visible sector in a truly realistic string compactification.

\subsection{DSB from D-branes at singularities}
\label{dsb}

An interesting spinoff in the study of D-branes at singularities has been
the realization of gauge theories whose non-perturbative dynamics removes the
supersymmetric vacuum 
\cite{Berenstein:2005xa,Franco:2005zu,Bertolini:2005di}. The prototypical 
example is provided by the gauge theory on a set of fractional branes on 
the $dP_1$ singularity. Moreover, the same behaviour is found in many 
other examples, and can in fact be argued to be generic. Nevertheless let 
us concentrate on the $dP_1$ case for concreteness (and for future 
application to our main examples).

The general gauge theory for D3-branes at a $dP_1$ singularity has been 
described in Section \ref{dimer}. Consider the particular situation where
\beqa
N_4=0 \quad, \quad N_1=M \quad,\quad N_2=2M \quad, \quad N_3=3M,
\label{dsbrank}
\eeqa
which corresponds to an anomaly-free, and hence consistent, choice. 
Recalling that the $U(1)$ gauge factors are massive (see 
footnote \ref{uunos}), the gauge group is $SU(3M)\times SU(2M)\times SU(M)$.
The superpotential reads
\beqa
W & = &  \, X_{23}X_{31}Y_{12}\, -\, X_{23} Y_{31} X_{12}
\label{dp1supo}
\eeqa
In addition there is the field $Z_{12}$, decoupled at this level.
As discussed in \cite{Berenstein:2005xa,Franco:2005zu,Bertolini:2005di} 
(see \cite{Intriligator:2005aw} for a detailed discussion), in the regime 
where the $SU(3M)$ dynamics dominates this gauge factor confines and 
develops an Affleck-Dine-Seiberg (ADS) superpotential for its mesons 
$M_{21} =  X_{23}X_{31}$, $M_{21}'= X_{23} Y_{31}$. The complete
superpotential is
\beqa
W & = &  M_{21}Y_{12}\, -\, M_{21}' X_{12}\, + \,
M \, \left( \, \frac{\Lambda_3^{7M}}{\det{\cal M}}\,\right)^{\frac 1M}
\label{unfldp1}
\eeqa
where ${\cal M}=(M_{21};M_{21}')$ is the mesonic $2M\times 2M$ matrix.
The theory has no supersymmetric vacuum since the F-term conditions for 
$X_{12}$, $Y_{12}$ send $M_{21}, M_{21}'\to 0$, and then the F-terms 
conditions for $M_{21}$, $M_{21}'$ send $X_{12},Y_{12}\to \infty$.
In fact, assuming canonical Kahler potential for the matter fields, one 
easily shows there is a runaway behaviour towards this minimum `at 
infinity' \cite{Franco:2005zu,Intriligator:2005aw}. The runaway direction 
is parametrized by the gauge-invariant dibaryonic operator
\beqa
\epsilon^{a_1\ldots a_{2M}} \epsilon^{b_1\ldots b_M} \epsilon^{c_1\ldots c_M}
(X_{12})_{b_1 a_1} \ldots (X_{12})_{b_M a_M} (Y_{12})_{c_1 a_{M+1}} \ldots
(Y_{12})_{c_M a_{2M}}
\eeqa
As mentioned above, this pattern is generic for a large class of systems 
of D-branes at singularities. Namely, for the so-called DSB fractional 
branes \cite{Franco:2005zu}, see \cite{Brini:2006ej} for the gauge theory 
analysis in a large set of examples.

It is useful to mention an equivalent viewpoint on the runaway
\cite{Franco:2005zu}. The $U(1)$
gauge factors can be maintained in the gauge theory, as long as one 
consistently includes the $B\wedge F$ couplings (and its supersymmetry 
related coupling of the NSNS scalar $\phi$ partner of $B$ as a 
Fayet-Iliopoulos (FI) term) in the dynamics, see footnote \ref{uunos}. 
Considering the FI $\phi$ for the linear combination $Q_1-Q_2$ of the 
$U(1)$'s in $U(M)\times U(2M)$, there is a non-trivial D-term constraint 
for fixed FI $\phi$, roughly of the form
\beqa
V_D\, =\, (\, |X_{12}|^2+|Y_{12}|^2 + \phi\, )^2
\eeqa
From this viewpoint, at fixed values of $\phi$ the D-term for the $U(1)$ 
lifts the runaway direction and leads to a non-supersymmetric minimum. In 
the complete theory, however, the FI term is actually a dynamical field 
$\phi$, which can decrease the vacuum energy to arbitrarily low values by 
relaxing to infinity. Hence the runaway behaviour is recovered, now in 
terms of the closed string mode $\phi$
\footnote{The relation between FI terms and baryonic operators is familiar
from several brane realization of gauge theories, starting with 
\cite{Elitzur:1997fh}. It would be desirable to have a precise map for 
systems with fractional branes.}.

This behaviour is interesting, but in principle it would seem of little 
phenomenological interest as a mechanism for supersymmetry breaking. 
However, it has recently been shown in \cite{Franco:2006es} that upon a 
small modification, the above class of theories (in particular the 
$dP_1$ theory) contain supersymmetry-breaking local minima, which are 
metastable and long-lived, since they are separated from the runaway 
behaviour at infinity by a large potential barrier
\footnote{Actually, since the field $Z_{12}$ is decoupled, the potential 
is invariant along the $Z_{12}$ direction, thus leading to a meta-stable 
supersymmetry breaking `valley' of vacua. The fate of this accidental 
left-over flat direction remains an open question.}.
The modification is a remarkably simple generalization of the proposal in 
\cite{Intriligator:2006dd} for SYM theories. It is provided by the  
introduction of massive flavours, with masses much smaller than the 
dynamical scale of the gauge theory. The additional flavours can be easily 
incorporated by the introduction of D7-branes in the system of D3-branes 
at singularities. We refer the reader to \cite{Franco:2006es} for details 
on the string construction and the gauge theory analysis of this
theory \footnote{As discussed in \cite{Franco:2006es}, the simplest
  realization of supersymmetry breaking minima in string theory would
  be via the introduction of D3- and D7-branes in a conifold
  singularity. However, in order to discuss the two possible
  realizations of supersymmetry breaking (stopping the runaway via
  moduli stabilization, and including D7-branes to get local minima)
  on equal footing, we center on constructions involving theories with
  DSB branes, like $dP_1$.}.

\begin{figure}
\begin{center}
\epsfxsize=12cm
\hspace*{0in}\vspace*{.2in}
\epsffile{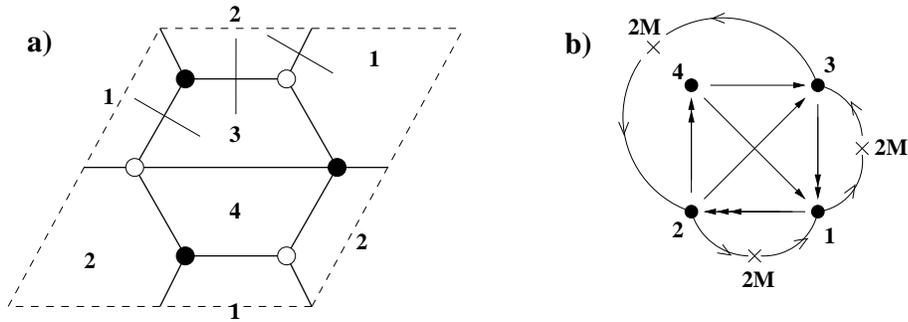}
\caption{\small (a) Dimer diagram for a configuration of D3- and D7-branes 
in the $dP_1$ singularity leading to a gauge theory with meta-stable 
supersymmetry breaking vacua. (b) Extended quiver diagram for the theory.}
\label{flavdp1}
\end{center}
\end{figure}

In Figure \ref{flavdp1} we show the extended dimer diagram corresponding 
to the system of D3- and D7-branes (with the rank assignment 
(\ref{dsbrank}) and $2M$ D7-branes). Other possible choices of D7-branes 
can in principle be similarly considered. In coming Sections we will use 
this configuration as a basic model of a sector leading to dynamical 
supersymmetry breaking (in its local non-supersymmetric minimum). 

An alternative possibility to obtain stable non-supersymmetric minima from 
DSB branes, already mentioned in \cite{Diaconescu:2005pc} is the following. 
As mentioned above, the runaway behaviour can be regarded as a non-trivial 
potential for a certain Kahler modulus of the singularity. In global 
compactifications, it is possible that there are other 
sources of potential for these moduli, which could presumably stabilize 
its runaway (for instance non-perturbative contributions arising from 
euclidean D3-brane instantons). This is however difficult to verify in 
concrete models including realistic sectors etc. Moreover, the properties 
of such local minima (including its very existence) would be strongly 
sensitive to the details of the global compactification. This goes against 
our strategy to attempt the construction of a visible plus DSB sector 
with no UV sensitivity.

In other words, one can rephrase the above by saying that in our 
specific local models, which are UV insensitive by construction, there are no 
other sources of potential for the Kahler moduli involved in
the runaway. Hence, the above proposal to modify the gauge theory by
adding slightly massive flavours is a UV independent way to generate 
supersymmetry breaking minima in these gauge theories, and a natural one
to be implemented in local models.

\subsection{Local CY models with several singularities}
\label{split}

\subsubsection*{Geometrical construction from partial resolution}

In this last subsection we would like to describe the construction of the
geometries of our interest, namely local Calabi-Yau varieties with two 
isolated singularities, and the gauge theories for D-branes placed on them. This is 
based on tools developed in \cite{Garcia-Etxebarria:2006aq}.

\begin{figure}
\begin{center}
\epsfxsize=12cm
\hspace*{0in}\vspace*{.2in}
\epsffile{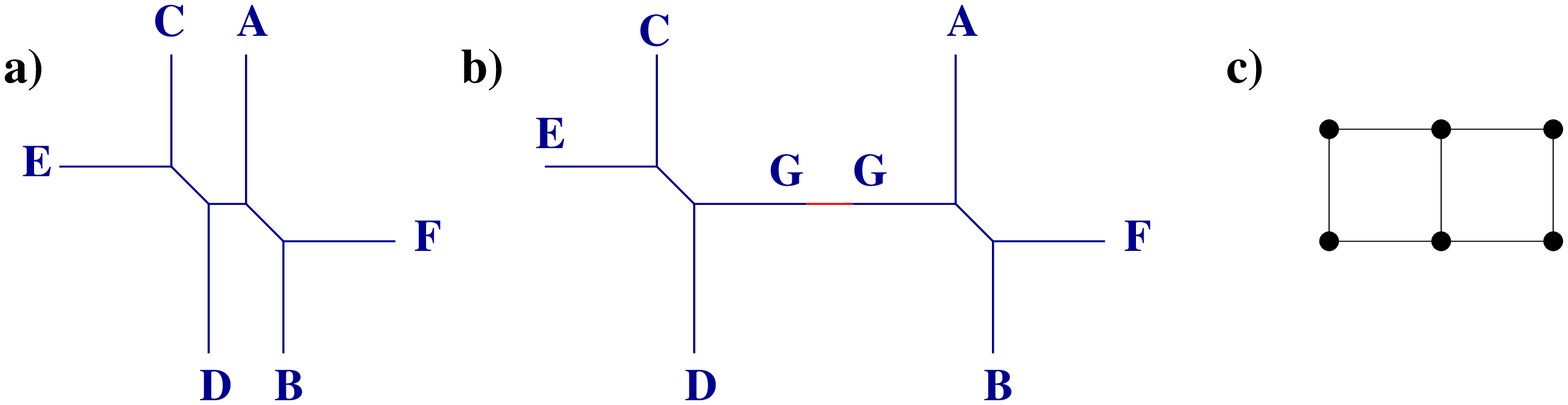}
\caption{\small (a) The web diagram for the double conifold singularity 
$xy=s^2w^2$. (b) The partial resolution to a geometry with two conifold 
singularities. (c) Description in terms of the toric diagram.}
\label{dconisplit}
\end{center}
\end{figure}

As mentioned above, non-compact toric Calabi-Yaus can be characterized 
using web diagrams. In this language, the construction of local CYs with two
isolated singularities is straightforward, by the procedure of partial 
resolution. We start with a web diagram describing a geometry with a single
singular point, namely all finite segments and faces are collapsed to a point
at which all external legs converge. Now we grow one finite size segment
\footnote{In some cases there exist partial resolutions involving simultaneous
growing of several parallel segments. They lead to geometries with two
singularities which are not isolated, but rather joined by a curve of 
singularities. We will not be interested in this case.} out 
of such a point. The web diagram now has two internal vertices at which 
external legs converge. This implies that the geometry now has two singular
points, separated by a distance controlled by the Kahler modulus of the 
2-cycle corresponding to the finite segment. The ideas are better 
illustrated using a concrete example. Hence, we consider an example 
studied in \cite{Garcia-Etxebarria:2006aq}, namely the splitting of 
the so-called double conifold singularity (studied in
\cite{Uranga:1998vf,Aganagic:1999fe}) to 
two conifold singularities. The web diagram for this singularity is shown 
in Figure \ref{dconisplit}a, and a partial resolution is illustrated in 
Figure \ref{dconisplit}b. 

The geometry of the two daughter singularities can be studied by
considering all legs entering the corresponding vertex (including the finite
size segment). Namely, by breaking the finite segment we obtain two daughter
web diagrams which describe the local geometry around the two daughter 
singularities \footnote{Note that in some cases one of the singularities may 
actually be a smooth space. We will not be interested in such cases.}.
This is manifest in Figure \ref{dconisplit}b, where, upon breaking the 
elongated segment (by removing the red piece in Figure \ref{dconisplit}b) 
we are left with two web diagrams describing the two conifold 
singularities in the left-over geometry. Notice that the original and 
final singularities are simpler to recognize if one keeps track of the
collapsed finite segments, by showing them with a small size, as we do in 
all our discussions. Recall however that the singularities are 
obtained when such finite pieces have zero size.

Notice that this process can be easily inverted. If one is interested in 
constructing a local CY with two isolated singularities of specified type,
one simply needs to consider combining their web diagrams into a larger one
by joining one external leg of each diagram into one finite size segment
\footnote{Notice that in doing so, we have the freedom to perform an 
$SL(2,\IZ)$ transformation on the web diagrams to facilitate the gluing.}.
This will be useful in the construction of geometries in Section \ref{basic}.

Finally let us mention that the partial resolution, when regarded in terms 
of the mirror Riemann surface $\Sigma$, simply corresponds to elongating
a tube. By pinching this tube (or elongating it infinitely) one obtains 
two daughter Riemann surfaces that describe the mirror of the two daughter 
singularities.

\subsubsection{Effect on D-branes}
\label{effect}
We would like to describe the effect of the above partial resolutions on 
systems of D-branes at the original singularity. This is most easily
determined using the language of dimer diagrams in previous Sections.

For D3-branes, this has been systematically described in 
\cite{Garcia-Etxebarria:2006aq}, which we review in what follows. 
Consider the gauge theory on D3-branes at the initial singularity.
Following the proposal in \cite{Morrison:1998cs}, the partial resolution 
corresponds to giving a vev to a closed string 
Kahler modulus. This couples as a FI term for the $U(1)$ gauge fields in
the D3-brane gauge theory, which therefore forces a set of bi-fundamental
multiplets to acquire a vev, breaking the gauge group by a Higgs mechanism.
After the Higgs mechanism, one recovers two gauge sectors, which are 
decoupled at the level of massless states, and which represent the gauge
theory on D3-branes at the two daughter singularities. The massive states
correspond to the massive open strings stretching between D3-branes at
different singularities. In the case where the two gauge sectors correspond
to the visible and supersymmetry breaking sectors, the massive modes are
the messengers of supersymmetry breaking.

The above discussion can be made very explicit using the dimer diagrams.
Consider the dimer diagram describing the gauge theory on D3-branes
at the initial singularity. As discussed above, the zig-zag paths of the
dimer diagram correspond to the external legs of the initial web diagram.
When the partial resolution is carried out, the web diagram splits into
two daughter web diagrams. The dimer diagrams of D3-branes at e.g. the first
daughter singularity can be obtained by using the initial zig-zag paths 
that correspond to external legs of the first daughter web diagram. To these
one should add a new zig-zag path corresponding to the external leg of the
daughter web diagram that arises from the finite size segment in the initial
web diagram. The resulting set of zig-zag paths determines the subset of
edges of the initial dimer diagram that survive in the first daughter dimer
diagram. Similarly for the second.

\begin{figure}
\begin{center}
\epsfxsize=10cm
\hspace*{0in}\vspace*{.2in}
\epsffile{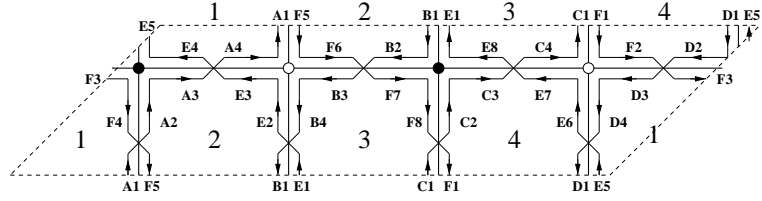}
\caption{\small Zig-zag paths for the dimer diagram of the
  double conifold. The path names agree with the names of the legs in the 
  web diagram in Figure \ref{dconisplit}a, and the numbers label the
  different gauge factors.}
\label{dconipath}
\end{center}
\end{figure}

\begin{figure}
\begin{center}
\epsfxsize=10cm
\hspace*{0in}\vspace*{.2in}
\epsffile{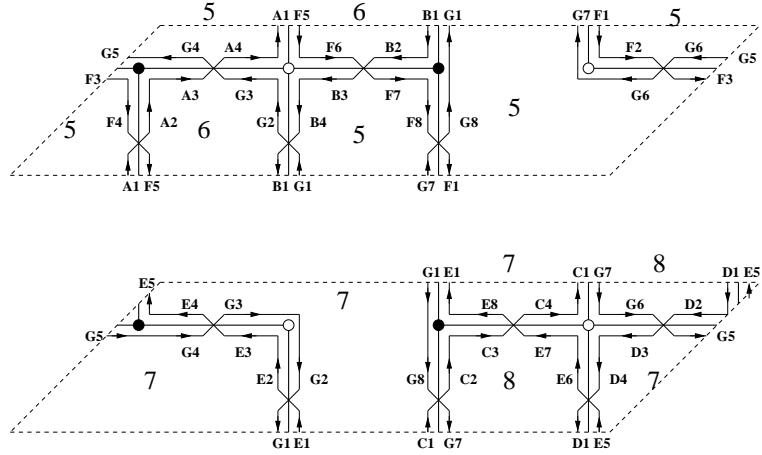}
\caption{\small Zig-zag paths corresponding to the two daughter theories,
  in the splitting of the double conifold singularity to two conifold
  singularities, with the corresponding dimers shown as thick lines. The
  numbers label the different gauge groups.}
\label{dconidaug1}
\end{center}
\end{figure}

To illustrate this, consider the example of the partial resolution of the 
double conifold to two conifolds \cite{Garcia-Etxebarria:2006aq}. The 
dimer diagram for the double conifold and its zig-zag paths are shown in 
Figure \ref{dconipath} (the $(p,q)$ homology charge of the paths 
correspond to the $(p,q)$ label of the legs in the web diagram in 
Figure \ref{dconisplit}a, modulo an overall $SL(2,\IZ)$ transformation). The 
partial resolution splits the web diagram into two daughter web 
diagrams, involving the legs A, F, B, G and C, D, E, G, respectively, 
see Figure \ref{dconisplit}b. The resolved geometry thus contains two 
singularities, at which two subsets of the original set of D3-branes are 
located. The dimer diagram of the gauge theory on D3-branes at the first 
daughter singularity is obtained by keeping the edges involved in the 
paths A, F, B of the original dimer diagram (with the new path G passing 
through edges crossed only once by paths in the set A, F, B).
Similarly for the second daughter singularity. The two daughter dimer 
diagrams are shown in Figure \ref{dconidaug1}. They correspond
(upon integrating out chiral multiplets with mass terms in the 
superpotential due to bi-valent nodes in the dimer diagram, see footnote 
\ref{bivalent}) to the dimer diagrams of conifold theories, in agreement with 
the effect on the geometry.

As discussed in \cite{Garcia-Etxebarria:2006aq}, the specific pattern of 
edges that survives in the different daughter dimer diagrams determines 
the specific vevs acquired by the bi-fundamental multiplets in the 
Higgsing of the initial gauge theory. Specifically,
let us denote edges of type 1 those disappearing in the 
second daughter diagram, of type 2 those disappearing in the first, and of 
type 3 those present in both (namely, those through which the path G 
passes). Denoting the corresponding  bifundamental vevs by  $\Phi_1$, 
$\Phi_2$, $\Phi_3$ the pattern of vevs for those fields in the partial 
resolution Higgs mechanism is
\beqa
\Phi_3 \, = \, \pmatrix{0 & 0 \cr 0 & 0} \quad ; \quad
\Phi_2 \, =\, \pmatrix{v\, \id_{N_1} & 0 \cr 0 & 0} \quad ; \quad
\Phi_1 \, =\, \pmatrix{0 & 0 \cr 0 & v\, \id_{N_2} }
\label{vevansatz}
\eeqa
where $N_1$, $N_2$ denote the number of D3-branes at the first and 
second daughter singularity. These vevs are flat with respect to the 
F-terms and non-abelian D-terms. Their deviations from $U(1)$ D-flatness 
is compensated by the FI terms controlled by the closed string modes 
carrying out the geometric blow-up.

The Higgs mechanism interpretation allows one to obtain the spectrum of 
massive states in the partially resolved geometry (namely the massive open 
strings stretching between D3-branes at different singularities) by 
starting with the initial gauge theory and computing the spectrum of 
multiplets becoming massive in the Higgs mechanism. The computation 
reduces to some dimer diagram gymnastics, and is described in Appendix 
\ref{massive} (this can be considered a new appendix of 
\cite{Garcia-Etxebarria:2006aq}). The result can be summarized as follows:

$\bullet$ {\bf 1} For each edge which disappears in the $i^{th}$ daughter dimer 
diagram, there is a massive vector multiplet in the adjoint of the 
$U(N_i)$ gauge factor corresponding to that location (i.e. that
arising from the diagonal of the gauge factors of the two faces the edge 
used to separate in the initial theory). 

$\bullet$ {\bf 2} For each face in the original dimer diagram, we obtain 
two massive vector multiplets in the bi-fundamental $(N_1,\ov{N_2})$ and 
its conjugate, of the gauge factors at the corresponding location.

$\bullet$ {\bf 3} For each edge present in both daughter dimer diagrams, 
there is 
one $(N_1,\ov{N_2})$ chiral multiplet in the corresponding bi-fundamental 
representation (i.e. charged under faces separated by the edge) becoming 
massive. The dimer diagram ensures that globally, these types chiral 
multiplets pair up consistently to form massive scalar multiplets.

$\bullet$ {\bf 4} Finally, if the daughter dimer diagrams contain bi-valent 
nodes 
(nodes with two edges) the corresponding edges each describe a massive scalar 
multiplet in the bi-fundamental of the two faces they separate.

Let us illustrate this with an example. For 
instance, the partial resolution of the double conifold to two 
conifolds is given by the following spectrum:

{\bf Vector multiplets in the adjoint:} There are two edges of type 1,
both giving rise to massive vector multiplets in the adjoint of the 
gauge factor 7 (see Figure \ref{dconidaug1}). Similarly, the two edges
of type 2 give massive vector multiplets in the adjoint of 5.

{\bf Vectors in the bifundamental}
We obtain massive vectors in the representations
\beqa
(5, \overline 7) \, +\, (6, \overline 7) \, +\, (5, \overline 7) \, +\, 
(5, \overline 8) \, + \, {\rm{c.c.}}
\eeqa

{\bf Scalar multiplets}
One finds the following spectrum of massive scalar multiplets:
\beqa
2\,(5, \overline 7)\, +\, (6, \overline 7)\, +\, (5, \overline 8)
\eeqa
Other examples are worked out similarly. A more complicated resolution 
will be described in section \ref{example}.

A last important point is that partial resolutions may be obstructed
whenever the
initial configuration contains fractional branes wrapped on the collapsed
cycle that acquires finite size in the partial resolution. Fractional branes
not of this kind can be regarded as fractional branes of the daughter
singularities (since they wrap cycles which remain collapsed to zero size
even after the partial resolution). These rules are manifest using the 
description of D3-branes as 1-cycles in the mirror Riemann surface $\Sigma$.
Fractional branes corresponding to 1-cycles stretching along the tube that
elongates in the partial resolution obstruct it. On the other 
hand, 1-cycles not stretching along the tube correspond to 1-cycles of 
the daughter Riemann surfaces, and hence define fractional branes of the 
daughter singularities. In our applications we are interested in this last 
kind of situation, hence the partial resolutions we consider are not 
obstructed. The generalization of the Higgs mechanism interpretation to 
situations with fractional branes is straightforward.

\subsubsection*{Including D7-branes}

The effect of partial resolutions on D7-branes was not described in 
\cite{Garcia-Etxebarria:2006aq}, but the discussion can be carried out 
using the description in Section \ref{dimer}. 

Recall the representation of D7-branes as segments across an edge in the 
dimer diagram (leading to 37, 73 states coupling to the corresponding 33
bi-fundamental in the D3-brane gauge theory). Let us consider the possible
D7-branes in the parent dimer diagram, and consider their fate in a partial 
resolution. This is essentially determined by the behaviour of the edge in
this process:

- A D7-brane across an edge which survives only in the first daughter 
dimer diagram, survives as a D7-brane passing through the first daughter 
singularity. It corresponds to the D7-branes of the daughter singularity
naturally associated to the corresponding edge in the daughter dimer (namely
leading to 73, 73 states coupling to the 33 bi-fundamental of the
corresponding gauge sector in the daughter theory).

- Similarly for D7-branes across edges surviving only in the second
daughter dimer diagram.

- Finally, a D7-brane across an edge that survives in both
daughter dimer diagrams corresponds to a D7-brane passing through both 
daughter singularities. 

\begin{figure}
\begin{center}
\centering
\leavevmode
\epsfxsize= 4.5in
\epsfbox{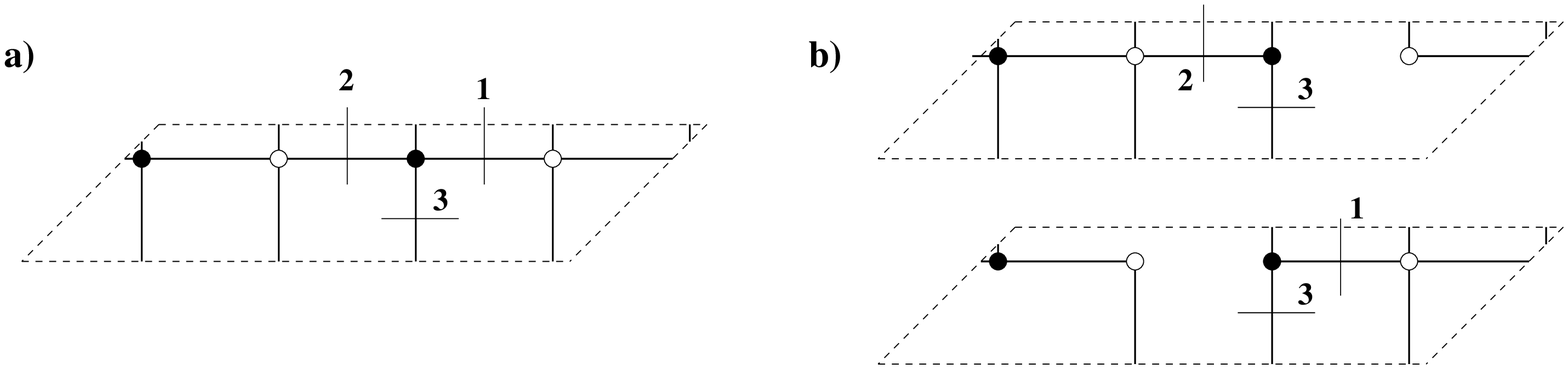}
\end{center}
\caption[]{\small The fate of different D7-branes in a partial resolution.
D7-branes associated to edges of type 1 resp. 2  become D7-branes absent 
in the first resp. second dimer diagram, hence passing through the second 
resp. first daughter singularity. For type 3 edges, the D7-branes remains 
in both daughter dimer diagrams, hence passes through both daughter 
singularities.}
\label{dconid7s}
\end{figure}

In Figure \ref{dconid7s} we provide examples of these 
possibilities in the partial resolution of the double conifold to two 
conifolds.

The rules are easily justified by considering the picture of D7-branes
as 1-cycles in the mirror Riemann surfaces, stretching between two punctures. 
Recall also that a D7-brane is naturally associated to a dimer diagram edge
(in the sense that the corresponding 33, 37, 73 states couple) over which
the two zig-zag paths associated to the punctures overlap. From this it
follows that D7-branes stretching between two punctures remaining in e.g. the 
first daughter Riemann surface, descend to D7-branes of the first daughter 
singularity. They are naturally associated to edges which survive only on
the first daughter dimer diagram (since the two zig-zag paths correspond 
to punctures surviving in the first daughter Riemann surface). Similarly for
D7-branes represented by 1-cycles stretching between punctures remaining in 
the second Riemann surface. The last possibility is a D7-brane represented 
by a 1-cycle stretching between punctures ending up in different daughter 
Riemann surfaces. Since it passes through the elongated tube in the 
partial resolution, it leads to two D7-branes in the two daughter 
theories.

The above description nicely fits with the field theory description in 
terms of Higgsing. A D7-brane leads to D3-D7 states with couplings 
37-73-33 with the 33 bi-fundamental associated to the edge across which 
the D7-brane segment stretches. If the edge disappears from e.g. the first
daughter dimer diagram, the corresponding 33 entries get a vev and give 
masses to the open string states stretching between the D7's and the first 
stack of D3-branes. On the other hand, open string states stretching 
between the D7's and the second stack of D3-branes remain massless, hence 
the D7-branes passes through the second daughter singularity, and can be 
represented as a segment in the second daughter dimer diagram (across the 
same edge). Similarly for edges disappearing in the second dimer diagram. 
Finally, for edges appearing in both daughter dimer diagrams, the 33 
bi-fundamentals get no vev, so all D3-D7 open string states remain 
massless, showing that the D7-brane passes through both daughter 
singularities. From this discussion it is clear that the rule to obtain 
the massive set of multiplets from D3-D7 open string states is:

$\bullet$ {\bf 5} For each D7-brane passing through an edge of type 1 
(resp. type 2) there is a massive scalar multiplet in the fundamental
representation of the $U(N_2)$ (resp $U(N_1)$) gauge factor
corresponding to the resulting recombined face. For $N_{D7}$ across
such an edge the massive multiplets transform as $(N_{D3},{\ov N}_{D7})$.

\section{Basic strategy and some examples}
\label{basic}

As announced, we plan to construct systems of D-branes at a local CY with 
two singular points, leading to two chiral gauge theories describing the
visible and supersymmetry breaking sectors. The system reproduces a model 
of GMSB in the regime where the distance between the D-brane stacks is 
smaller than the string scale. In most of the paper, we consider the case 
where the distance is controlled by a Kahler modulus (on which we center 
in this paper). The system is thus most efficiently described as a slight 
partial resolution of a worse singularity, as those we have discussed. 
Correspondingly, the 
complete gauge system (two sectors plus messengers) are fully encoded on 
the gauge theory of D3-branes at this worse singularity, in the presence 
of non-trivial FI terms (triggering the above described Higgsings). 
In this case, the distance between the singularities is classically a flat 
direction (however possibly getting a non-trivial potential upon 
supersymmetry breaking). In order to avoid dealing with this issue, which 
we leave as an open question, we assume that all Kahler moduli have been 
stabilized.

The case where the distance is controlled by a complex structure parameter 
admits an analogous interpretation. The 
geometry is most efficiently described as a slight complex deformation of 
a worse singularity. The complete gauge system is encoded in the gauge 
theory on D3-branes at the latter, in the presence of fractional branes, 
triggering the complex deformation via a geometric transition. An important 
difference with the previous situation is that the distance between 
singularities is dynamically fixed in terms of the amount of fractional 
branes triggering the geometric transition. The idea is sketched in 
Section \ref{deform}, and although realistic models are possible,
they tend to be involved and we do not present any explicit example.

It is important to realize that, although the idea of using CY 
geometries with several singularities is general, it is most practically 
implemented in toric geometries, on which we center in most of the paper.

\subsection{A simple example}
\label{example}

Let us consider one simple example of the above strategy. We would like to 
consider a non-compact Calabi-Yau with two singularities, with their local 
structures being that of a complex cone over $dP_0$, and a complex cone 
over $dP_1$ respectively. We would like to locate D3-branes at each of 
these singularities, so as to obtain two gauge sectors, which are 
decoupled at the level of massless states (although massive open strings 
stretched between the two stacks provide a massive messenger sector). 

\begin{figure}
\begin{center}
\centering
\epsfysize=3.5cm
\leavevmode
\epsfbox{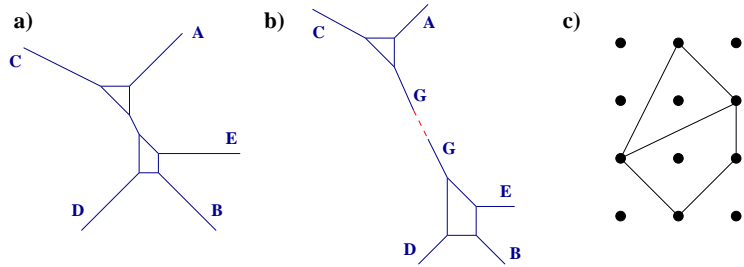}
\end{center}
\caption[]{\small (a) Web diagram for a local CY with $dP_0$ and  
$dP_1$ singularities, for generic sizes of all 2- and 4-cycles. (b) The 
two singularities are obtained when the cycles corresponding to the two 
finite faces shrink to zero size, while the leg G remains finite and 
controls the distance between the singularities. (c) Toric diagram for the 
geometry, with the partial resolution leading to the two separated 
singularities.}
\label{x31split}
\end{figure}

The simplest toric geometry realizing this is described by the web diagram
in Figure \ref{x31split}a. As usual, and for clarity, we have shown the 
geometry with all 2- and 4-cycles of finite size. The geometry of 
interest, with the two singularities is better represented by Figure 
\ref{x31split}b, more specifically when the two small faces in the web 
diagram are collapsed to zero size. The two singularities are described by 
the two sets of blue legs. The finite leg G with the dashed red piece on it 
describes the 2-cycle which controls the distance between the two 
singularities, and thus the mass scale of the messenger sector. 

Regarding Figure \ref{x31split}b as preceding Figure \ref{x31split}a 
illustrates a simple algorithm to construct local Calabi-Yau geometries 
containing several singularities. One simply 
considers the web diagrams for the different daughter singularities, and 
glues them together by combining external legs of the daughter 
web diagrams into finite size legs (which is always possible by 
using the $SL(2,\IZ)$ freedom in defining each of the daughter web 
diagrams) \footnote{In doing this, some additional external legs may 
cross, implying that they are actually internal legs in the complete 
diagram, see Section \ref{possib} for some such examples.}.

D3-branes at the $dP_0$ singularity can provide a toy model of the MSSM. 
For the time being, we can consider e.g. 3 D3-branes (without fractional 
branes) at the $dP_0$ singularity, so that the gauge theory content is
\beqa
{\rm Vector:} \quad & \quad  U(3)_{1'}\times U(3)_{2'}\times U(3)_{3'} 
\nonumber\\
{\rm Chiral:} & 3\, [\, (3,\ov{3},1)\, +\, (1,3,\ov{3})\, +\, (\ov{3},1,3)
\, ]
\eeqa
and there is a superpotential coupling (\ref{dp0supo}). In fact, this is 
one example (very similar to \cite{Willenbrock:2003ca}) of the 
so-called trinification models extending the MSSM.

Similarly, D-branes at the $dP_1$ singularity can provide a toy model of a 
sector with dynamical supersymmetry breaking. Specifically, we consider 
introducing $M$ fractional D-branes at the $dP_1$ singularity, so that the 
gauge theory is precisely that studied in Section \ref{dsb}, namely
\beqa
{\rm Vector:} \quad & \quad  U(3M)\times U(2M)\times U(M) \nonumber\\
{\rm Chiral:} & (3M,\ov{2M},1)\, +\, 3(1,2M,{\ov M})\,
+\, 2(\ov{M},1,M)
\eeqa
Recall there is a superpotential coupling (\ref{dp1supo}), and that the 
$U(1)$'s are actually massive due to their couplings to closed 
string modes, see footnote \ref{uunos}.

Modulo the runaway issue in the $dP_1$ theory (to be fixed via the 
stabilization of Kahler moduli, or by the addition of massive flavors as
in the next Section), this is a simple  
configuration realizing gauge mediated supersymmetry breaking in a local 
setup. In particular, it is a very tractable example of a theory similar 
to those introduced in \cite{Diaconescu:2005pc}. Moreover, it 
has the advantage that new ingredients can be easily added to improve 
its properties, so that new variants are easily implemented. For instance 
it is straightforward to introduce D7-branes to turn the runaway $dP_1$ 
sector into the flavoured $dP_1$ theory with a local supersymmetry 
breaking minimum discussed in Section \ref{dsb} (see Section 
\ref{complete}).

A more fundamental advantage is that it is possible to describe explicitly 
the physics of this theory when the scale of mediation is small compared 
with the string scale. Namely, when the distance between the two 
singularities (and hence of the two D-brane stacks) is shorter than the 
string length. Since this distance is controlled by a K\"ahler parameter, 
classical geometry is not a good approximation. The real dynamics is 
captured by including the messenger sector, which is far lighter than the 
cutoff scale (string scale), in the effective theory. Namely, by 
considering the complete field theory obtained when the two singularities 
coalesce, with the two singularities arising from a small partial 
resolution. Equivalently, with the gauge symmetry slightly broken by Higgs 
expectation values (induced by the FI terms from the closed string vevs 
blowing up the singularity) as described in Section \ref{split}. Moreover, 
as emphasized there, one can keep track of the multiplets becoming massive 
in the Higgs mechanism to obtain an explicit description of the massive 
messenger sector. So this is one of the few frameworks where such a spectrum 
is actually computable.

\begin{figure}
\begin{center}
\centering
\epsfysize=3.5cm
\leavevmode
\epsfbox{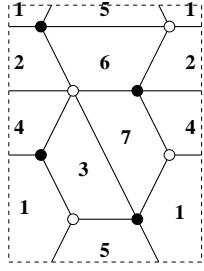}
\end{center}
\caption[]{\small The dimer diagram for the $X^{3,1}$ theory.}
\label{x31dimer}
\end{figure}

As should be clear, we thus need to consider the geometry whose web 
diagram is shown in \ref{x31split}a, in the limit where {\em all} finite 
pieces collapse to zero size, and there is a single singularity, and 
study the gauge theory on D3-branes (and fractional branes) at such a 
singularity.
For a general toric singularity, one can use general techniques to obtain
such field theories. Happily, this task has already been carried out in 
our case. The geometry of interest is a particular example $X^{3,1}$ in 
the infinite family of geometries $X^{p,q}$ introduced in 
\cite{Hanany:2005hq}. The dimer diagrams for the gauge theory on D3-branes 
at these geometries have been determined in \cite{Franco:2005rj}, and for 
the $X^{3,1}$ it is shown in Figure \ref{x31dimer} (with a relabeling of 
faces with respect to the latter reference).

\begin{figure}
\centering
\psfrag{dP0}{$dP_0$}
\psfrag{dP1}{$dP_1$}
\includegraphics[scale=0.50]{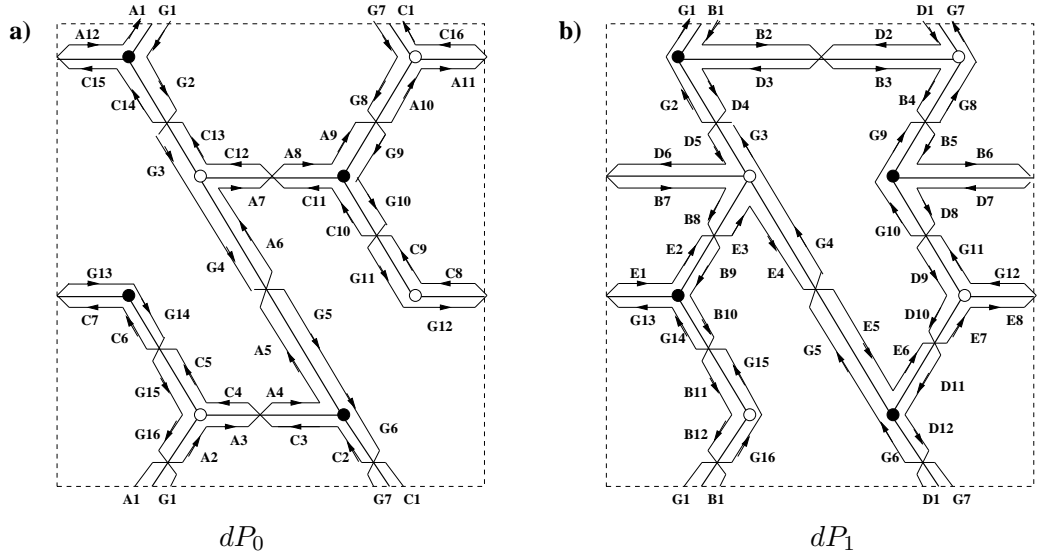}
\caption{\small Daughter dimer diagrams obtained in the partial resolution 
of $X^{3,1}$ to a geometry with $dP_0$ and $dP_1$ singularities. The dimer 
diagrams indeed describe the gauge theories of D3-branes at these two 
singularities.} 
\label{x31daughter}
\end{figure}

\begin{figure}
\centering
\psfrag{1}{$1'$}
\psfrag{2}{$2'$}
\psfrag{3}{$3'$}
\includegraphics[scale=0.50]{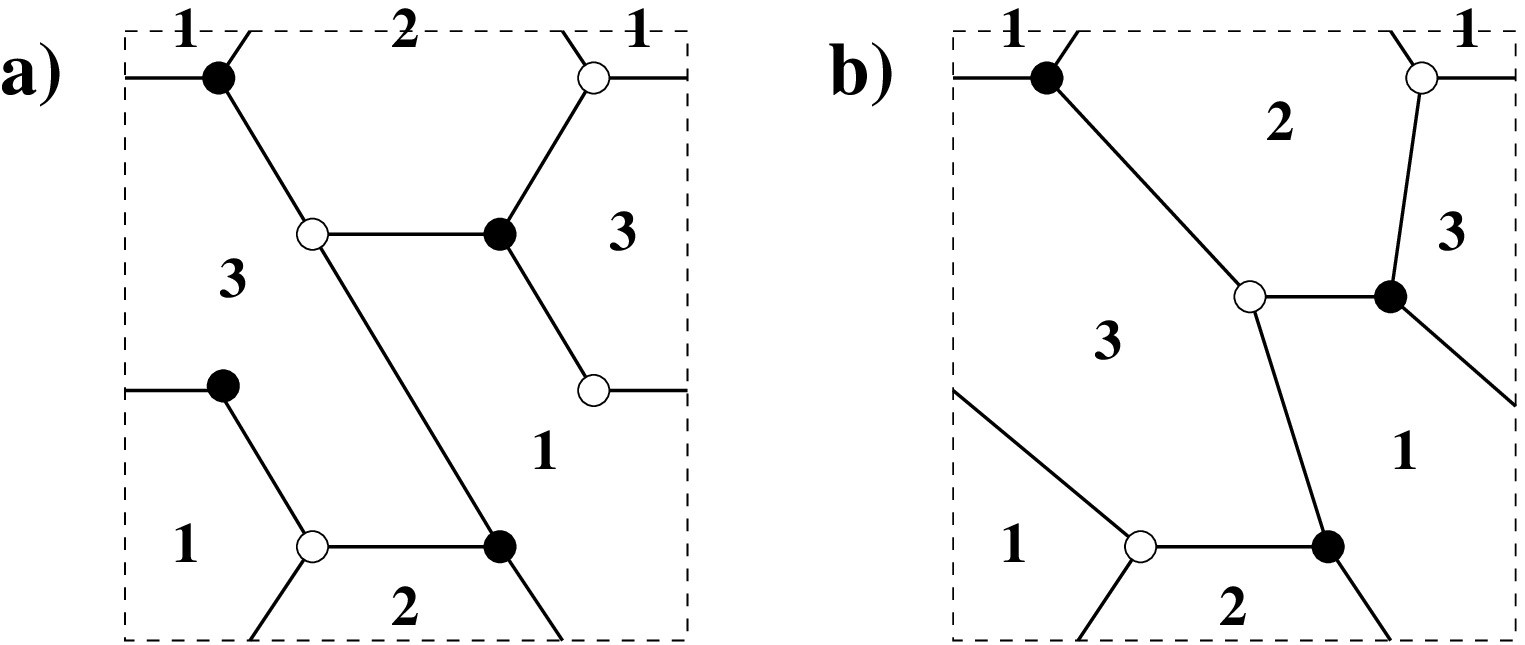}
\caption[]{\small (a) The dimer diagram in Figure \ref{x31daughter}a. 
(b) Upon integrating out matter massive due to bi-valent nodes one obtains 
a dimer diagram corresponding to the $dP_0$ theory.}
\label{x31massage1}
\end{figure}

\begin{figure}
\begin{center}
\centering
\epsfysize=3.5cm
\leavevmode
\epsfbox{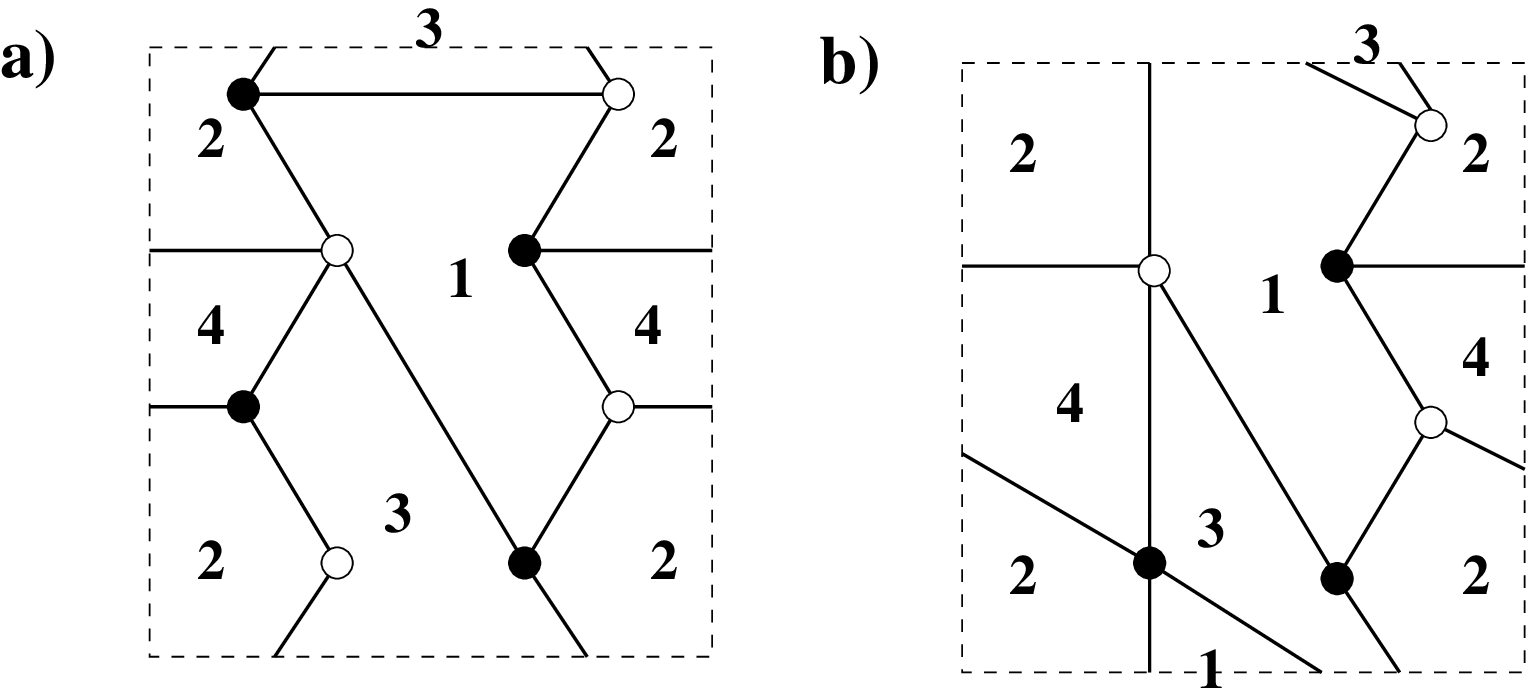}
\end{center}
\caption[]{\small (a) The dimer diagram in Figure \ref{x31daughter}b. 
(b) Upon integrating out matter massive due to bi-valent nodes one obtains 
a dimer diagram corresponding to the $dP_1$ theory.}
\label{x31massage2}
\end{figure}

The partial resolution of the $X^{3,1}$ singularity to a 
geometry with $dP_0$ and $dP_1$ singularities is a Higgs mechanism which 
can be studied as in Section \ref{split}. Namely the splitting of the web 
diagram into two daughter web diagrams, as in Figure \ref{x31split}b, 
leads to two sets of paths (A, C, G, and B, D, E, G) which define the 
daughter dimer diagrams of the D3-branes at the two daughter 
singularities. The paths and the resulting daughter diagrams are shown in 
Figure \ref{x31daughter}a, b. In fact, after integrating out 
matter with mass couplings in the superpotential (due to bi-valent nodes), 
one can show they correspond to the dimer diagrams of D3-branes at the 
$dP_0$ and $dP_1$ singularities, respectively. This is shown in Figure
\ref{x31massage1} for the $dP_0$ case and in Figure \ref{x31massage2} for 
the $dP_1$ case.

\begin{figure}[!htp]
\centering
\psfrag{dP0}{$dP_0$}
\psfrag{dP1}{$dP_1$}
\includegraphics[scale=0.50]{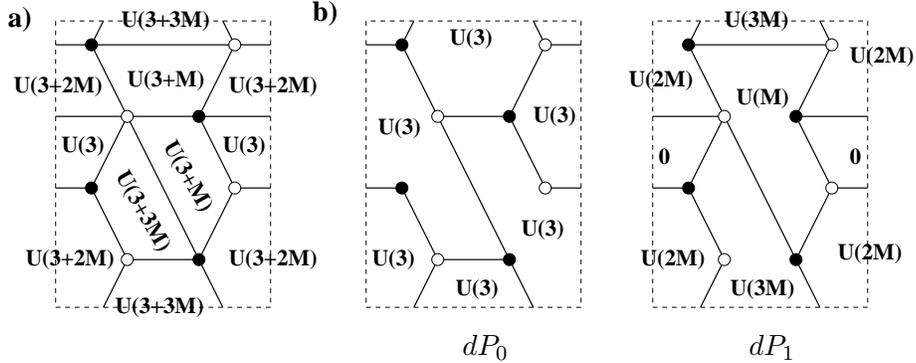}
\caption{\small Rank assignment for the gauge theories (a) when the two 
singularities are collapsed into an $X^{3,1}$ singular point, (b) for the 
two gauge sectors corresponding to D3-branes at the isolated $dP_0$ and 
$dP_1$ singularities obtained after partial resolution.} 
\label{x31fract}
\end{figure}

Given this general framework, we can be more specific about the choice of 
D3-brane structure we are considering. We consider the particular case
\beqa
N_1=3+2M \quad , \quad
N_2=3+2M \quad , \quad
N_3=3+3M \quad , \quad
N_4=3 \nonumber \\
N_5=3+3M \quad , \quad
N_6=3+M \quad , \quad
N_7=3+M \quad , \quad
\eeqa
Namely 3 regular D3-branes and one fractional D-brane. The dimer diagram 
for the original $X^{3,1}$ theory is shown in Figure \ref{x31fract}a, and 
the dimer diagrams for the two stacks of branes after partial 
resolution are shown in Figure \ref{x31fract}b. Notice that the total 
rank on each region of the original dimer diagram is equal to the sum 
of the ranks in the corresponding regions in the daughter dimer diagrams. 
This is the condition for consistent partial resolution in the presence of 
fractional branes determined in \cite{Garcia-Etxebarria:2006aq}. Following 
the rank assignment in Figure \ref{x31fract}b through the process in 
Figure \ref{x31massage2}, it is easy to see that the fractional brane 
descends to a fractional brane of the daughter $dP_1$ singularity. 
Namely, using the notation in Figures \ref{dp0dimer} and 
\ref{dp1dimer}, the rank assignments in each daughter gauge theory are
\beqa
& dP_0 & : \quad N_1'=N_2'=N_3'=3 \nonumber \\
& dP_1 & : \quad N_1=M\, ,\, N_2=2M\, ,\, N_3=3M
\eeqa
So we easily identify the two gauge theory sectors described at the 
beginning of this Section.

As discussed in Section \ref{split}, the dimer diagram allows to 
easily read off the bi-fundamental vevs leading to this Higgsing, and to 
obtain the massive spectrum of mediators. We will use the notation of
Figures \ref{x31massage1} and \ref{x31fract} for the gauge groups in
$dP_0$ and $dP_1$ respectively, and we denote fundamental
representations for the group $1'$ by $1'$ and fundamentals of $U(2M)$
by $2M$ (similarly for antifundamentals).

{\bf Vector multiplets in the adjoint:} There are 7 massive vector 
multiplets in adjoint representations, coming from the 7 edges of type 2 
(associated to the fields $X_{56}$, $X_{24}$, $X_{43}$, $X_{71}$) or type 
1 (associated to $X_{21}$, $X_{76}$ and $X_{56}$). They lead to massive 
vector multiplets in the representation
\beqa
{\rm Ad}_{1'}\, +\, {\rm Ad}_{2'}\, +\, 2\,{\rm Ad}_{3'}\, +\, 
{\rm Ad}_{M}\, +\, {\rm Ad}_{2M}\, +\, \,{\rm Ad}_{3M} 
\eeqa

{\bf Vectors in bi-fundamentals:} There is one such massive vector 
multiplet for each face in the original gauge group. They transform 
in the representation
\beqa
(1', \overline {2M}) \, +\, (3', \overline {2M}) \, +\, (3', \overline 
{3M}) \, + \nonumber \\
(2', \overline {3M}) \, +\, (2', \overline M) \, +\, (1', \overline M)
+ {\rm c.c.}.
\eeqa
Note that face 4 of the $X^{3,1}$ dimer does not contribute, as the
corresponding gauge factor in the $dP_1$ dimer has rank 0 with our
choice of fractional brane.

{\bf Scalar multiplets}
Using the rules described in Section \ref{effect}, one finds the following
spectrum of scalar multiplets:
\beqa
(3',\overline {2M})\, +\, 2\, (1', \overline {3M})\, +\,
2\, (2', \overline {2M}) \, +\, 2\, (3', \overline M)
\eeqa
It would be interesting to compute the effects of supersymmetry breaking 
in models of this kind. We leave this for future work.

\subsection{A more complete construction}
\label{complete}

As already mentioned, one additional advantage of the present setup is its 
flexibility. For instance, maintaining the same geometry, it is extremely 
simple to describe variants of the theory in the previous Section, by changing 
the D-brane configuration. We illustrate this by building a version 
with a more realistic visible sector, and an improved supersymmetry breaking 
sector (in that it is independent of the stabilization of Kahler moduli 
needed to prevent the runaway of the $dP_1$ theory in the previous 
section).

\begin{figure}[!htp]
\centering
\psfrag{dP0}{\LARGE{$dP_0$}}
\psfrag{dP1}{\LARGE{$dP_1$}}
\includegraphics[scale=0.60]{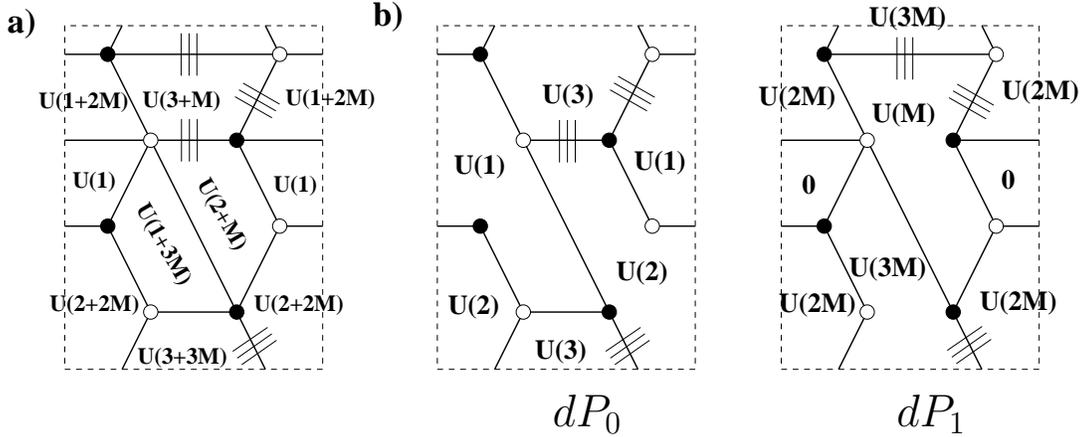}
\caption{\small \small Gauge theories with D7-branes at (a)  $X^{3,1}$, (b) $dP_0$ and $dP_1$ singularities obtained after partial resolution.} 

\label{x31fract_D7}
\end{figure}

This can be done by adding D7-branes in $X^{3,1}$. Figure 
\ref{x31fract_D7}a shows the D7-branes present in the original
$X^{3,1}$ singularity as well as the rank assignment arising from
fractional branes. We have not shown the N regular D3-branes
present in $X^{3,1}$ and which only survive in the $dP_1$ sub-dimer.
As stated in Section \ref{effect}, D7-branes crossing edges of type
1(2) only survive in sub-dimer 1(2), whereas D7-branes crossing edges
of type 3 appear in both sub-dimers. Thus, after resolution we obtain
the dimers in Figure \ref{x31fract_D7}b which correspond to the
quivers shown in Figures  \ref{dp0mssm} and \ref{flavdp1} with the
rank of the flavour gauge groups in $dP_1$ equal to 3. Also, the
condition for the supersymmetry breaking sector $dP_1$ to contain
supersymmetry breaking local minima which are metastable and
long-lived imposes  $M=2$ in Figure \ref{x31fract_D7}.

\section{Some additional possibilities}
\label{possib}

In this Section we describe some generalizations and other model building 
possibilities, which, although they involve more complicated geometries, 
lead to interesting or novel features.

\subsection{Flavour universal supersymmetry breaking for $\IC^3/\IZ_3$}

The $X^{3,1}$ (or $dP_0+dP_1$) model studied in Sections \ref{example} 
and \ref{complete}, has an important drawback from the viewpoint of the
phenomenology of supersymmetry breaking. Namely, the complete geometry 
treats the three families in an asymmetric way, eventually resulting in a 
lack of universality in the soft terms, in particular the squark masses, 
in conflict with known constraints in flavour physics.

The root of the problem is the following. The different families on
D3-brane models at singularities are associated to the three complex 
directions of the transverse space. Hence in the $\IC^3/\IZ_3$ 
singularity the three 
families are treated symmetrically \footnote{In models with D7-branes, 
they have to be introduced in a way that maintains this, but it can be 
easily arranged, see \cite{Aldazabal:2000sa}.}. This symmetry appears in 
the web 
diagram as a cyclic rotation of the diagram (this is manifest when the 
diagrams are shown in a slightly tilted way, see coming pictures), 
implemented by the order-3 $SL(2,\IZ)$ action 
\beqa
\pmatrix{-1 & -1 \cr 1 & 0}
\label{z3gen}
\eeqa 
acting 
on the $(p,q)$ labels. The symmetry of the configuration is however not 
preserved by the complete $X^{3,1}$ geometry, as is manifest in the web 
diagram. Upon partial resolution, one recovers the $\IC^3/\IZ_3$ 
singularity, and the symmetry of the different families at the level of 
the massless spectrum and its interactions, but not in the interactions of 
the different families with the massive messenger sector.

Understanding of this problem leads to a natural solution. One should 
enforce the symmetry between complex planes in the complete geometry. 
Namely, the complete web diagram should be invariant under the action of 
(\ref{z3gen}). This is easily achieved by construction: the $X^{3,1}$ 
geometry can be regarded as obtained by adding a $dP_1$ web diagram to the 
$dP_0$ web diagram, along a specific external leg of the latter. The 
choice of this special leg breaks the symmetry between the complex planes.
Therefore, in order to preserve the symmetry, the same operation must be 
carried out in all external legs of the $dP_0$ web diagram. Namely we end 
up with a geometry obtained by adding three $dP_1$ web diagrams along the 
three legs of the $dP_0$ diagram. 

\begin{figure}
\begin{center}
\includegraphics[scale=0.3]{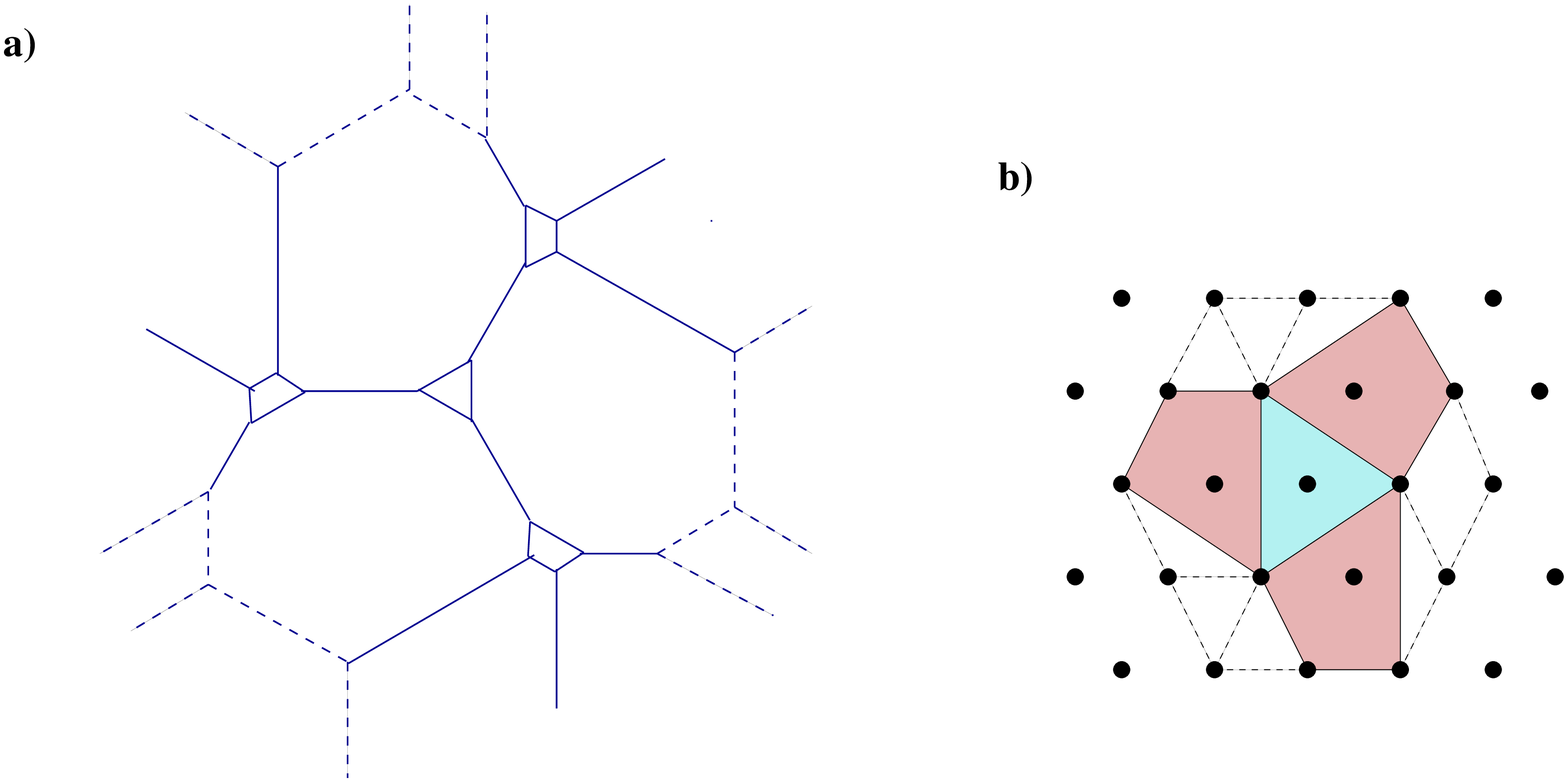}
\end{center}
\caption[]{\small The web diagram (a) and toric diagram (b) for a 
singularity admitting a partial resolution with one $dP_0$ singularity 
(in blue) and three (symmetrically distributed) $dP_1$ singularities (in 
red).}
\label{z3sym}
\end{figure}

In Figure \ref{z3sym} we show the 
web diagram and toric diagram of a possible resulting geometry 
\footnote{When the $dP_1$ web diagrams are added to the $dP_0$ one, some 
of the external legs of the former cross. This simply means that they 
are actually internal legs of the complete web diagram. The final 
external legs stem from junctions of the crossing legs. Figure 
\ref{z3sym} illustrates one possible choice of such junctions, leading to 
a relatively simple geometry.}. We have shown the diagrams slightly tilted
in order to make the $\IZ_3$ symmetry manifest. Despite the complicated 
appearance of the diagrams, they are in principle tractable, since the 
complete geometry corresponds to an orbifold of the complex cone over 
$dP_3$, for which the dimer diagram and gauge theory data are 
easily computable. The result is however not particularly illuminating, 
and we skip its discussion. 

Nevertheless, the idea is that upon partial resolution, which can 
be systematically analyzed, we obtain a $dP_0$ 
gauge theory, describing a visible sector, coupled in a flavour symmetric 
way to three $dP_1$ supersymmetry breaking sectors. Choosing the dynamical 
scale of the latter gauge sectors equal (in line with the symmetry we try 
to preserve), the soft terms induced in the $dP_0$ sector arise 
symmetrically for the three families \footnote{Of course it is an 
interesting question to determine the extent to which this symmetry 
constraints other properties of the model, like its Yukawa couplings. We 
leave this kind of analysis as an open question.}. 

\subsection{${\bf \Delta_{27}}$ models}

Although we have focused on toric geometries, it should be clear that the 
main idea of using local CY geometries with several D-brane sectors is 
more general. A simple generalization would be to consider non-toric 
geometries, for instance local CY geometries containing a non-abelian 
orbifold singularity. In fact, this kind of generalization has an 
immediate application, since D-branes at the non-abelian orbifold 
singularity $\IC^3/\Delta_{27}$ have been suggested to lead to 
semirealistic visible sectors \cite{Aldazabal:2000sa,Berenstein:2001nk,
Verlinde:2005jr}.

In fact, we can use our tools to construct e.g. a local CY geometry with 
one sector of D3-branes at a ${\bf \Delta_{27}}$ singularity, and a 
supersymmetry breaking sector of D-branes at a $dP_1$ singularity. 
The $\IC^3/\Delta_{27}$ singularity is obtained by modding out $\IC^3$ 
(parametrized by complex coordinates $z_1,z_2,z_3$) by the actions
\beqa
\theta & : & (z_1,z_2,z_3) \to (\alpha z_1,\alpha^{-1}z_2,z_3) \nonumber 
\\
\omega & : & (z_1,z_2,z_3) \to ( z_1,\alpha z_2,\alpha^{-1} z_3) \nonumber 
\\
\sigma & : & (z_1,z_2,z_3) \to (z_2,z_3,z_1)
\eeqa
with $\alpha=e^{2\pi i/3}$.

The quotient by the subgroup generated by $\theta$, $\omega$ defines a 
$\IC^3/(\IZ_3\times \IZ_3)$ geometry, which is a toric geometry. Hence 
$\IC^3/\Delta_{27}$ can be regarded as a quotient of this geometry by the
$\IZ_3$ action generated by $\sigma$, which cyclically permutes the 
three complex planes. Similarly, a geometry containing a ${\bf 
\Delta_{27}}$ 
singularity and e.g. a $dP_1$ singularity can be regarded as a $\IZ_3$ 
quotient of a geometry containing a $\IC^3/(\IZ_3\times \IZ_3)$ 
singularity and {\em three} $dP_1$ singularities with a distribution 
invariant under cyclic permutation of the complex coordinates. 

\begin{figure}
\begin{center}
\includegraphics[scale=0.3]{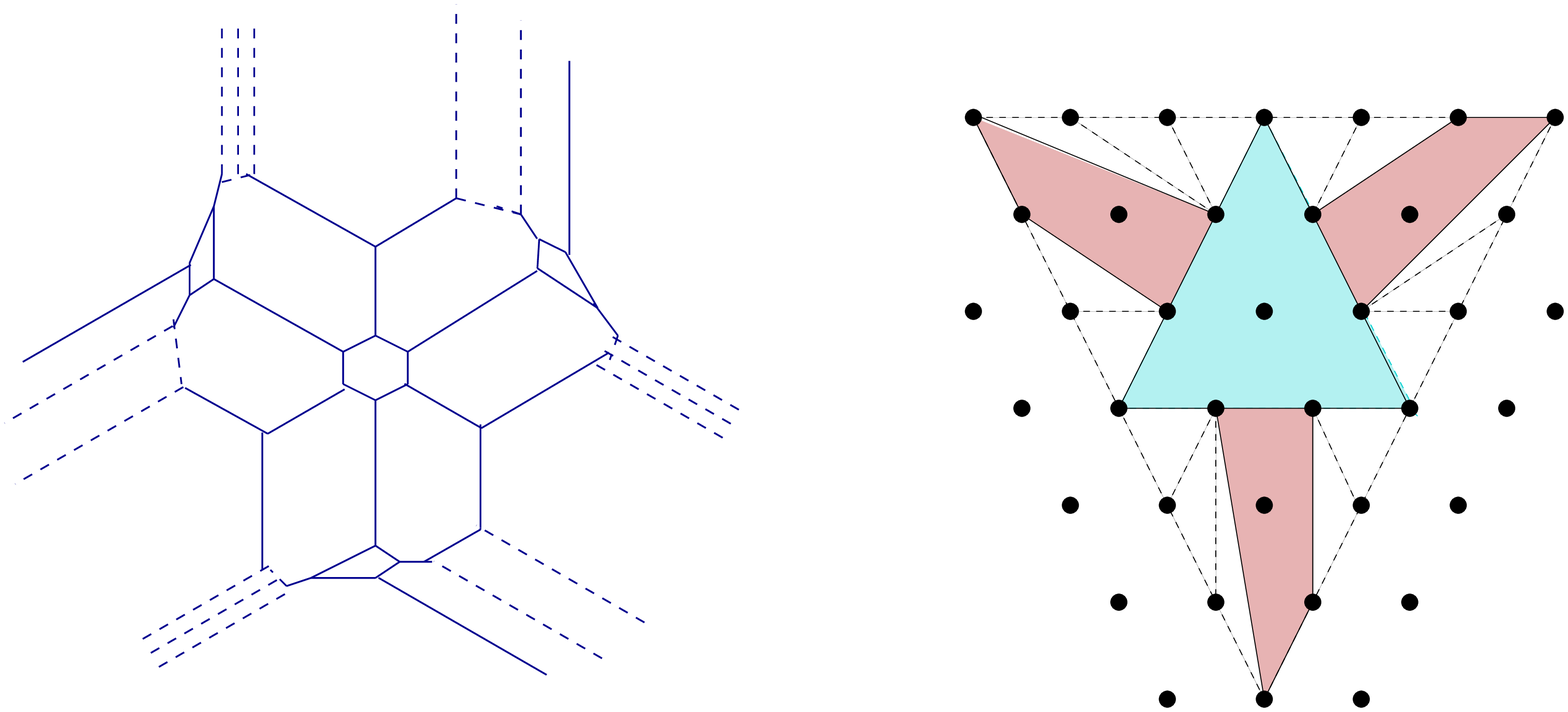}
\end{center}
\caption[]{\small The parent toric geometry for the
  $\IC^3/\Delta_{27}$ singularity with a supersymmetry breaking sector
  (given by $dP_1$) coupling in an universal way to the visible sector.}
\label{Delta27}
\end{figure}

The construction of these `parent' geometries and the corresponding 
gauge theories is easy, and identical at the technical level to the previous 
section. Namely, we consider the web diagram of the $\IC^3/(\IZ_3\times 
\IZ_3)$ singularity, and add the web diagrams of three $dP_1$ 
singularities in a way that preserves the $\IZ_3$ symmetry of the diagram.
The web and toric diagrams of a possible complete geometry are shown in 
Figure \ref{Delta27}. Despite the complicated appearance of the 
diagram, it is closely related to a $\IC^3/(\IZ_6\times \IZ_6)$ 
singularity, so it is tractable in principle. Notice however the different 
spirit 
of the construction as compared to previous section, in that here we are  
ultimately interested in quotienting by the permutation symmetry among 
the complex planes, in order to generate the ${\bf \Delta_{27}}$ 
singularity. 
Once a parent geometry is constructed, and the gauge theory identified, 
the quotient gauge theory can be obtained by using techniques in 
\cite{Uranga:1998vf}.

We hope this example suffices to illustrate the use of our ideas in 
somewhat more general contexts.

\subsection{Complex deformed CY's with several singularities}
\label{deform}

In this paper we have described the construction of configurations of 
D-branes at different singularities in a local CY geometry, obtained by 
partial resolution. This has the advantage of allowing for a simple 
computation of the messenger sector. On the other hand, it requires an 
additional discussion of the stabilization of distance between singularities 
(via some mechanism of Kahler moduli stabilization, whose description is 
not completely clear in the local model).

We would like to briefly mention an alternative proposal, based on CY 
geometries where the structure of singularities arises after complex 
deformation. 
As in the previous situation, the construction of a geometry 
containing e.g. two isolated singularities of the desired kind can be 
systematically carried out, by combining the corresponding web diagrams. 
Specifically, a toric singularity admits a complex deformation to a 
geometry with two daughter singularities if the external legs of the web 
diagram of the parent singularity can be separated in two subsets, 
corresponding to the external legs of the web diagrams of the daughter 
singularities. This criterion, used in \cite{Aganagic:2001ug,Franco:2005fd}
in the physical context, dovetails the mathematical description in 
\cite{altmann1,altmann2,altmann3}. 

The complex deformation setup has the advantage that the modulus 
controlling the distance between the singularities is a complex structure 
modulus, which can be stabilized using 3-form fluxes. One may interpret 
this as a source of UV sensitivity. However, the 
complex structure deformation can be entirely described in terms of 
the gauge theory of the initial singularity, as the confining gauge 
dynamics of a set of fractional D-branes (the so-called deformation 
branes), in analogy with \cite{Klebanov:2000hb,Franco:2005fd}. Efficient 
tools to carry out this gauge theory analysis, and hence determine the 
effect of complex deformation on the D-brane sectors, have been introduced 
in \cite{Garcia-Etxebarria:2006aq}. From the viewpoint of the gauge 
theory, the distance between the final D-brane stacks is related to the 
strong dynamics scale of the deformation fractional branes, clearly 
showing that it is not a modulus of the configuration (more precisely, 
it still has a dependence on the string coupling, which is nevertheless 
not a local modulus, hence its stabilized value depends on the global 
structure). In fact, it is this gauge theory description, rather than the 
geometric one, which is reliable in the regime of interest where the 
distance between the singularities is smaller than the string scale.

Hence one can in principle describe the complete dynamics in terms of the 
gauge theory associated to D-branes at the singularity obtained by 
shrinking all cycles in the geometry. This is similar to what happened in 
the partial resolution setup. However, the splitting of 
this initial singularity into several is a strong coupling effect triggered 
by confinement of the deformation fractional branes. The low-energy 
dynamics after this confinement can be determined reliably, and leads to 
two decoupled sectors corresponding to D3-branes at the two daughter 
singularities. On the other hand, the messenger sector corresponds to the 
massive states of the confining theory (with mass determined by the strong 
dynamics scale, or the complex deformation parameter in geometric terms)
and cannot be reliably computed.

\begin{figure}
\begin{center}
\includegraphics[scale=0.3]{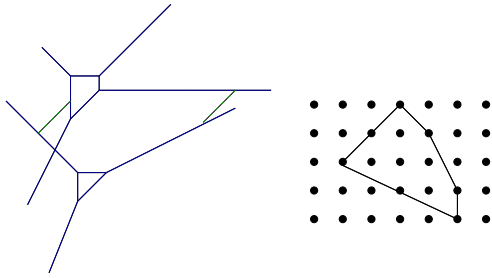}
\end{center}
\caption[]{\small Web diagram and toric diagram of a singularity admitting 
a complex deformation to a geometry with a $dP_0$ and a $dP_1$ 
singularities. For clarity we have shown the web diagram explicitly split 
in two sub-webs, namely after the complex deformation. The finite size 
3-cycle is shown as a dashed green segment.} 
\label{def}
\end{figure}

Since this setup lacks the computability of the partial resolution setup, 
we prefer to skip its detailed discussion, and simply mention one example 
of a singularity admitting a complex deformation to a geometry with 
a $dP_0$ and a $dP_1$ singularity. The relevant web diagram and toric 
diagrams are shown in Figure \ref{def}. The model building application of 
such 
configurations are very similar to those described in the partial 
resolution setup. The relevant gauge theory analysis to obtain the final 
two decoupled gauge theory sectors from the gauge theory of D3-branes at 
the parent singularity are provided in \cite{Garcia-Etxebarria:2006aq}, to 
which we refer the interested reader.

\section{Final comments}
\label{comments}

In this paper we have exploited tools to construct systems of D3-branes 
at CY geometries with several singularities in order to embed models of 
GMSB in string theory. The construction is simple and flexible and allows 
many generalizations. It would be interesting to develop tools to study 
the effects of supersymmetry breaking both on the visible sector, and on 
the geometry itself. In this latter respect, it would be interesting to 
determine the effects of supersymmetry breaking on the Kahler moduli which 
control the distance between the brane stacks (and which we have assumed 
to be stabilized at a high scale).

Notice that although very interesting, models of D3-branes at 
singularities are not the only possible realization of GMSB in local 
models. For instance, it would be interesting to develop local models 
where the D-branes in the visible and DSB sector are not necessarily of 
the same kind (parallel D3-branes in our case). One can e.g. imagine a 
local geometry with two sets of intersecting D-branes leading to two 
chiral sectors decoupled at the level of massless states. Such models 
would perhaps be more generic and rich, but simple problems in our 
setup, like the determination of the messenger sector, are probably 
difficult in these other situations.

We leave these and other issues as interesting open questions for the 
future.

\centerline{\bf Acknowledgements}

We thank S. Franco, L. E. Ib\'a\~nez for useful conversations . F.S. and 
I.G.-E. thank CERN TH for hospitality during completion of this project. 
A.M.U. thanks Tel Aviv University for hospitality, and M.~Gonz\'alez for
kind encouragement and  support. This work has been partially supported by
CICYT (Spain) under project FPA-2003-02877, and the RTN networks
MRTN-CT-2004-503369 `The Quest for Unification: Theory confronts
Experiment', and MRTN-CT-2004-005104 `Constituents, Fundamental Forces
and Symmetries of the Universe'. The research of F.S. is supported by
the Ministerio de Educaci\'on y Ciencia through an F.P.U grant. The
research of I.G.-E. is supported by the Gobierno Vasco PhD fellowship
program and the Marie Curie EST program.

\bigskip

\appendix

\section{Massive sector in partial resolutions}
\label{massive}

In this appendix we provide the derivation of the spectrum of states 
becoming massive in the partial resolution of a singularity into two. The 
derivation is based on the description in \cite{Garcia-Etxebarria:2006aq}, 
and can in fact be considered an additional appendix to that reference.

In a partial resolution, the dimer diagram leads to two daughter 
dimer diagrams. Denote F, E, V and F$_i$, E$_i$, V$_i$, $i=1,2$ the number 
of faces, edges and vertices in the initial and daughter diagrams. Recall 
they satisfy the Euler formulas $F-E+V=0$, $F_i-E_i+V_i=0$. Also, each 
daughter dimer diagram has the same vertices as the initial one, hence 
$V_i=V$. Finally, we denote $N_i$ the number of D3-branes at the $i^{th}$ 
daughter singularity, and $N=N_1+N_2$ the initial number.

The number of gauge bosons becoming massive in the Higgs mechanism 
associated to the partial resolution (namely $U(N)^F\rightarrow 
U(N_1)^{F_1}\times U(N_2)^{F_2}$) is
\beq
n_V\, =\, F(N_1+N_2)^2 - F_1(N_1)^2 - F_2(N_2)^2 = (F-F_1)N_1^2 + 
(F-F_2)N_2^2 + 2FN_1N_2
\eeq
Also, the number of chiral multiplets which become massive is
\beq
n_{ch}\, =\, E(N_1+N_2)^2 - E_1(N_1)^2 - E_2(N_2)^2 = (E-E_1)N_1^2 + 
(E-E_2)N_2^2 + 2EN_1N_2
\eeq
Of these latter, $n_V$ of them are eaten by the massless vector multiplets 
to lead to massive vector multiplets. Using the Euler formulas and 
$V_i=V$ we have 
\beq
(F-F_i)- (E-E_i) = F-E - (F_1-E_1) = 0 \; \rightarrow \; F-F_i = E-E_i
\eeq
Hence $(E-E_i)N_i^2$ chiral multiplets are eaten by the $(F-F_i)N_i^2$ 
vector multiplets, and similarly $2FN_1N_2$ chiral multiplets out of the 
$2FN_1N_2$ are eaten by the corresponding vector multiplets. The remaining 
chiral multiplets, which are $2(E-F)N_1N_2 = 2VN_1N_2$ in number, pair up 
into massive scalar multiplets via superpotential terms as we show below.

Now, let us try to specify how all the multiplets become massive. 
Consider first the $(F-F_1)N_1^2$ disappeared vector multiplets. The 
disappearance is due to the fact that some faces in the initial diagram 
recombine in the first daughter diagram. They do so because there are 
$(E-E_1)$ edges which have disappeared, due to the vev of the $N_1\times 
N_1$ block in the corresponding bi-fundamental. This shows that the 
$(F-F_1)N_1^2$ vector multiplets eat up the $(E-E_1)N_1^2$ chiral 
multiplets, leading to $F-F_1=E-E_1$ massive vector multiplets in the 
adjoint of the $U(N_1)$ gauge symmetry of the corresponding recombined 
face. Similarly for the $(F-F_2)N_2^2$ vector and chiral multiplets. This 
is rule number {\bf 1} in Section \ref{effect}.

\begin{figure}
\begin{center}
\includegraphics[scale=0.4]{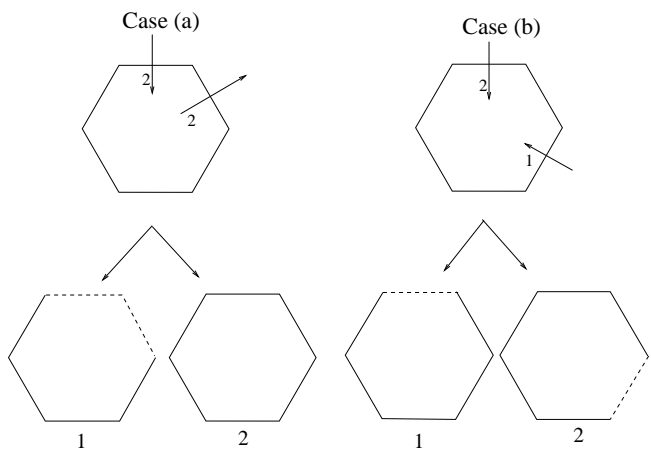}
\caption{Two possible configuration of edges for a face.}
\label{fig1}
\end{center}
\end{figure}

In order to understand the additional $2FN_1N_2$ disappeared vector 
multiplets, it is useful to have a more precise picture of how the edges 
of a face in the initial diagram can behave. Notice that for a given
face in the original dimer diagram, it is impossible that all edges are of 
type 3 (present in both sub-dimers). If all edges in a face would be of 
type 3, and given the fact that at each node there can only be two edges 
of type 3 (this will be proven later), then that face would correspond to 
a cycle on the Riemann surface wrapping the new puncture G coming 
from the resolution. However, since this cycle corresponds to a face in 
the dimer, its (p,q) charge would be zero, which is impossible. Thus 
every face has to have at least two edges which are not of type 3, so 
either two edges of the face are of type 1, i.e. disappear from sub-dimer 
2, (or two are of type 2) or one edge is of type 1 and another of type 2 
(see Figure \ref{fig1}). We denote these two cases (a) and (b)

\begin{figure}
\centering
\psfrag{a12}{$a_{12}$}
\psfrag{b12}{$b_{12}$}
\psfrag{A12}{$A_{12}$}
\psfrag{B12}{$B_{12}$}
\includegraphics[scale=0.40]{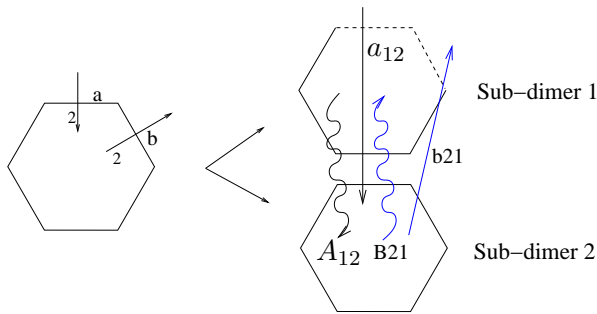}
\caption{\small 
Coupling between chiral and vector multiplets for the case when 
a face has two edges of the same type.} 
\label{case1_coupling}
\end{figure}

The $2FN_1N_2$ disappeared vector multiplets arise from 
open strings stretching between subdimers 1 and 2, at the same face 
location in both. They become massive by eating up chiral multiplets 
associated to open strings stretching between both sub-dimers, across 
disappeared edges. In case (a), the coupling occurs as shown in Figure 
\ref{case1_coupling}.
%

\begin{figure}
\centering
\psfrag{a12}{$a_{12}$}
\psfrag{b12}{$b_{12}$}
\psfrag{A12}{$A_{12}$}
\psfrag{B12}{$B_{12}$}
\includegraphics[scale=0.40]{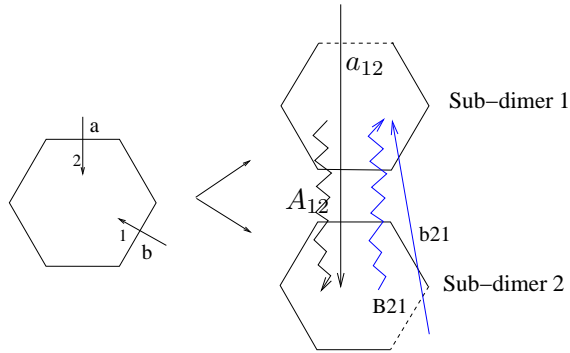}
\caption{\small Coupling between chiral and vector multiplets for the 
case when a face has one edge of each type.} 
\label{case2_coupling}
\end{figure}

%
The vector multiplets (shown as wavy arrows) $A_{12}$ and $B_{21}$ couple 
to the chiral multiplets $a_{12}$ and $b_{21}$ respectively (which 
stretch across edges $a$ and $b$ respectively).
In case (b), the coupling occurs as shown in Figure \ref{case2_coupling}.
%
%
The vector multiplets $A_{12}$ and $B_{21}$ couple to $a_{12}$ and 
$b_{21}$ respectively. $a_{12}$ and $b_{21}$ stretch across edges $a$ and 
$b$ respectively. This can be easily generalised to a face with an 
arbitrary assignation of edges. The above discussion shows that 
for each face in the original dimer diagram, we obtain 
two massive vector multiplets in the bi-fundamental $(N_1,\ov{N_2})$ and 
its conjugate, of the gauge factors at the corresponding location. This is 
rule number {\bf 2} in Section \ref{effect}

Let us now consider the remaining $2(E-F)N_1N_2=2VN_1N_2$ chiral 
multiplets. As we show, they become massive due to the $V$ superpotential 
terms. These chiral multiplets  arise from open strings stretching 
between the two dimer diagrams (with both orientations), across edges of 
type 3. The fact that each superpotential term leads to a mass for a chiral 
multiplet in the $(N_1,{\ov {N_2}})$ and $({\ov{N_1}},N_2)$ (of the faces 
separated by the corresponding edge) follows from the fact that each node 
has necessarily two edges of type 3. Namely, all fields in the 
superpotential term, except the two chiral multiplets, acquire vevs, 
leading to a mass term for the latter. Hence one recovers rule number 
{\bf 3} in Section \ref{effect}.

\begin{figure}
\begin{center}
\includegraphics[scale=0.4]{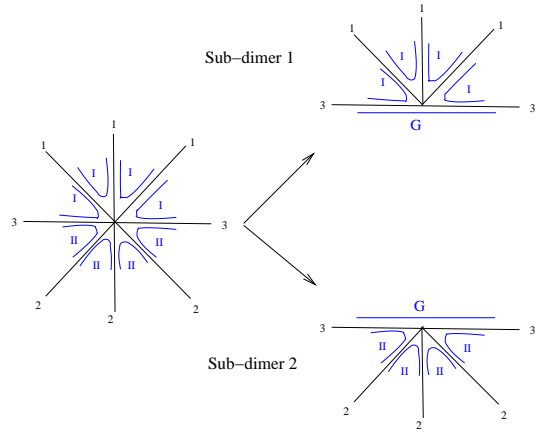}
\caption{Resolution for the case when only two edges at a given node
  are of type 3. G represents the new puncture which arises in the
  resolution}
\label{fig4}
\end{center}
\end{figure}

\begin{figure}
\begin{center}
\includegraphics[scale=0.4]{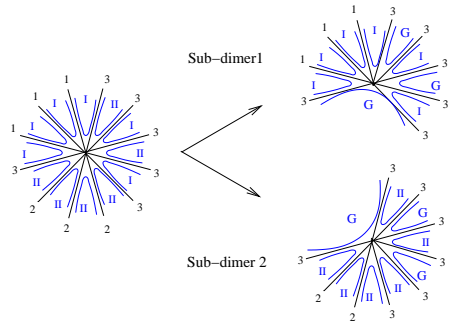}
\caption{Resolution for the case when more than two edges at a given
  node are of type 3. G represents the new puncture which arises in
  the resolution.}
\label{fig5}
\end{center}
\end{figure}

The property that each node necessarily has two edges of type 3 can be 
shown as follows. In a partial resolution, the zig-zag paths of the 
original dimer diagram are split in two sets I and II.
That is, the daughter dimer diagram 1 is obtained by removing the zig-zag 
paths II and adding the zig-zag path G which correspond to the new 
puncture. Similarly for dimer diagram 2, with the zigzag G being the 
same but with opposite orientation. Now, at each node, two edges of type 1 
and 2 have to be separated by at least one edge of type 3. A little 
thought shows that if there are more than two edges of type 3 at any 
given node, the zig-zags G in both subdimers cannot be the same. This is 
illustrated in Figures \ref{fig4} and \ref{fig5}. In the first 
Figure one sees that when only two edges of type 3 are present at a given 
node, then they separate the graph into two regions of type 1 and 2 
respectively. Now, in the daughter dimer diagram 1 (resp. 2) all edges of 
type 2 (resp. 1) are absent. Hence the zigzag G of the new puncture 
passes through the boundary of region 1 (resp. 2), consistently leading to 
the same G with opposite orientation in the two diagrams. The situation 
for a node with more than two edges of type 3 is shown in Figure 
\ref{fig5}. Since it clearly leads to paths G which are not the same in 
the two dimer diagrams, we conclude that such node structure is not 
possible.

One small subtlety is that for a given edge of type 3, there are actually 
two chiral multiplets becoming massive. These correspond to open strings 
stretching across this edge and going from the first daughter dimer 
diagram to the second and vice-versa. Each superpotential term pairs 
only one of these chiral multiplets (coupling it to only one of the 
chiral multiplets in the other adjacent type 3 node). And for 
a given edge of type 3, both modes acquire mass thanks to the two 
superpotential terms at the nodes of the edge.

\end{document}